\documentstyle[12pt]{report}
\input epsf
\begin{document}
%
\pagenumbering{roman}
\thispagestyle{empty}%
\null\vskip.5in%
\begin{center}                                                          
{\Large\uppercase{
The Supersymmetric Flavor Problem\\
and\\
Mu to E Plus Gamma\\}}                         
\end{center}                                                            
\vfill                                                                  
\begin{center}                                                          
\sc a dissertation\\                                            
submitted to the department of Physics\\                        
and the committee on graduate studies\\                         
of stanford university\\                                        
in partial fulfillment of the requirements\\                    
for the degree of\\                                             
doctor of philosophy                                            
\end{center}                                                            
\vfill                                                                  
\begin{center}                                                          
\rm By\\                                                        
David W. Sutter\\                                                      
June 1995\\  
hep-ph-9704390                                               
\end{center}\vskip.5in
\chapter*{Abstract}
The supersymmetric $SU(3)\times SU(2)\times U(1)$ theory with minimal particle content and general soft supersymmetry breaking terms has 110 physical parameters in its flavor sector: 30 masses, 39 mixing angles, and 41 phases.  These parameters contain many new measurable sources of flavor violations unless either the sparticle mass splittings and mixing angles are small or the sparticle masses are large.

In chapter~\ref{chap:prob} of this report, we discuss the origins of sparticle mass splittings and mixing angles in a theory explaining flavor, and we review the upper limits of sparticle masses that arise from naturalness among the electroweak breaking parameters.  By examining the flavor changing processes  $\mu \rightarrow e + \gamma$ and $K^0 - \bar{K}^0$ mixing, we show it is not possible to simultaneously satisfy the requirements of flavor differentiation among sparticles and naturalness in electroweak breaking.  This indicates that a crucial ingredient is missing from our understanding of the theory.  We discuss one possible solution in which the messengers that transmit supersymmetry breaking to ordinary particles are much lighter than $M_{\rm Planck}$.

In our analysis, we found the most important constraint was the process $\mu \rightarrow e + \gamma$.  Furthermore, this process is currently being experimentally investigated.  In spite of its importance, the complete branching ratio calculation has not yet been done.  In chapter~\ref{chap:calc} we present the full one-loop calculation for $\mu \rightarrow e + \gamma$.
%
\chapter*{Acknowledgements}
I would like to thank my advisor Savas Dimopoulos for all of the knowledge, guidance, and opportunities he gave me.  It was a joy working with him not only because of his insights, but also because of his sense of humor.  I would also like to give a special thanks to Lenny Susskind for the great deal I learned from him.  He had a way of making even the most complicated ideas simple.

Other professors who gave me special help me during my graduate student career are Bryan Lynn, Michael Peskin, Bob Laughlin, Andrei Linde, Renata Kallosh, and Bob Wagoner.  I would like to thank the members of my reading committee:  Professors Susskind, Peskin, Laughlin, Linde, and the chairmen Ben Andrews.

I owe a special thanks to Uri Sarid, with whom I held many important discussions and who gave generously of his time to help review the documents that make this thesis.  I also owe thanks to L{\'a}rus Thorlacius.

Many fellow students contributed to my stay at Stanford as a graduate student.  In particular, I would like to thank John Uglum, Jonathon Feng, Don Finnell Ya'el Shadmi, Amanda Peet, Arthur Mezhlumian, Andrew Lee, Michael Boulware, Matt Strassler, Misha Samoilov, and Dave Hochran.  I would also like to thank Massimo Giovannini who is a graduate student at Cern.

Thank you to the Physics department staff for all the support they gave me.  I am especially grateful of Karin Slinger, without whom I never could have survived as a graduate student.  She always made sure everything was going well for me.  Marcia Keating was particularly helpful both while I was at Cern, making sure everything was taken care of in the States, and while I was back home preparing to graduate.

I would like to thank the Theory Group at Cern for their hospitality during my year on leave there.  I would like to thank the TH secretariat, particularly Suzy Vascotto, for all of the help concerning working at Cern and surviving in a foreign country.

Finally, and most importantly, I would like to thank my parents, brother, and sister for the innumerable things they have done for me.
\tableofcontents

\listoffigures

\chapter*{Introduction}
\addcontentsline{toc}{chapter}{Introduction}
\pagenumbering{arabic}
\setcounter{page}{1}

The standard model has been a tremendous success.  It became firmly established with the discovery of the $W^{\pm}$ \cite{w} and $Z^0$ \cite{z} bosons, and has since been further confirmed by precision measurements at LEP and SLAC and by the discovery of the top quark \cite{top}.  This great success has however led to a problem for particle physics -- recent progress has been slow because there have not been any major unexpected experimental results.  There are indications, however, that this trend will come to an end.

For all the experimental successes of the standard model, there are still some theoretical difficulties.  The biggest problem is the stability of a light weak scale in the presence of quadratic divergences from physics at much higher scales \cite{quad}.  The solution to this problem is new physics near the weak scale, which may be seen at LEP~II, LHC, or the NLC.

Two major theories for this new physics have been developed:  technicolor \cite{tech} and the supersymmetric standard model \cite{dg}.  Technicolor theories face serious challenges by the small size of flavor changing neutral currents, whereas supersymmetry does not have such a large problem.  In the absence of supersymmetry breaking terms, the supersymmetric standard model has the same couplings as the standard model, making its phenomenology similar.   Furthermore, in addition to accommodating the experimental results of the standard model, the supersymmetric standard model also has experimental evidence that is not shared by the standard model: the successful weak mixing angle prediction in a SUSY GUT \cite{gutsin}.

When we do consider the supersymmetry breaking parameters, we have the potential for new phenomenology.  In the flavor sector alone, the supersymmetric standard model has 21 masses, 36 mixing angles, and 40 phases in addition to the standard model parameters.  These many parameters may appear to be a nightmare for interpreting experimental results, but they are a blessing for the future of particle physics.  These parameters convey information from high energy physics that could otherwise be beyond the reach of experiment.  The new flavor sector parameters could hold the answer to the two biggest mysteries of the supersymmetric standard model:  the structure of flavor and the breaking of supersymmetry.

Although we do not yet know any values for the new supersymmetric parameters, we can make conclusions based on the nonobservation of a supersymmetric contribution to flavor changing neutral currents.  In the first chapter of this report, which is an expanded version of reference~\cite{fp}, we examine this point and conclude that there must be some mechanism which prevents flavor violations from being large.

In the second chapter, we present the full one loop $\mu \rightarrow e + \gamma$ calculation.  The full calculation is necessary not only for precision evaluation of the supersymmetric parameters from $\mu \rightarrow e + \gamma$ decays, but it is also necessary for order of magnitude estimates in general regions of SUSY parameter space.

\chapter{The Supersymmetric Flavor Problem}
\label{chap:prob}

\section{Naturalness versus Flavor: A Conflict}

Nature is ambivalent about Flavor; Quark masses violate it significantly whereas neutral processes conserve it very accurately. This ambivalence leads to a conflict that has to be resolved in every theory. In the standard model it led to the GIM mechanism \cite{gim}. In SUSY-GUTS and the supersymmetric standard model \cite{dg} it led to the hypothesis that squarks and sleptons of the same color and charge have the same mass, independent of the generation that they belong to. We call this ``horizontal universality''. A stronger version of this hypothesis is that all squarks and sleptons have the same mass at  $M_{\rm GUT}$ \cite{dg}. This is called ``universality'' and is a fundamental ingredient of the minimal version of the Supersymmetric Standard Model (MSSM).  Universality ensures that the sparticle masses are isotropic in flavor space and thus do not cause any direct flavor violations. Flavor non-conservation  in the MSSM originates in the quark masses and is under control.

The hypotheses of universality or horizontal universality are difficult to implement in realistic theories. The reason is simple: the physics that splits particles also splits sparticles \cite{hkr}.  The degree to which this happens is the crucial question; the answer depends on the superpotential couplings and the nature of the supersymmetry breaking.  In the MSSM  and the minimal SUSY-GUT  \cite{dg} the interfamily sparticle splittings that are dynamically induced are adequately small. Such minimal theories leave fundamental questions unanswered and are unlikely to be the last word. In more ambitious theories addressing (even small parts of) the flavor structure of the standard model, the interfamily sparticle splittings are invariably large and cause unacceptable flavor violations unless the sparticle masses are heavy \cite{fpg}\cite{fp}.  However, heavy sparticles spoil naturalness, which was the original reason for low energy SUSY; it implies that parameters related to electroweak symmetry breaking must be tuned to high accuracy. Thus, generic theories addressing the problem of flavor conflict with naturalness.

In this paper, we first present the general supersymmetrized standard model.  Next we review sources of flavor dependence in the supersymmetry violating terms and the naturalness criterion that limits the masses of the new supersymmetric partners.  We then use the experimental constraints from the flavor changing processes $\mu \rightarrow e + \gamma$ and $K^0 - \bar{K}^0$ mixing to quantify the conflict between sparticle non-universality and naturalness, illustrating the need for a mechanism to suppress flavor changing processes.  Finally, we discuss one of the prospective suppression mechanisms.

\section{The Supersymmetrized Standard Model}
\label{sec:ssm}

The standard model has been an incredible success; it is consistent with virtually every particle physics experiment we can do.  Nonetheless, we believe it is an effective theory that is valid only up to energies just beyond those we can currently probe.  The reason for this is that the scalar fields responsible for the breaking of electroweak symmetry in the standard model have no protection from large loop corrections which should raise the weak scale to the scale of new physics \cite{quad}.

The most popular model for new physics beyond the Standard model is N=1 supersymmetry \cite{susy}.  In supersymmetry, the scalar particles are accompanied by fermions which conspire to cancel the dangerous quadratic divergences which spoil the naturalness of a light weak scale.  

To supersymmetrize the standard model, every field must be promoted to a superfield, which contains the original field plus a partner field which differs from the original by one half unit of spin.  Additionally, the supersymmetryized model requires an extra scalar doublet superfield because of the holomorphic condition for the superpotential.  The couplings of the model are the same as in the standard model with the exception of the Higgs quartic coupling, which is not an independent parameter.   Rather, it is a combination of the $SU(2)$ and $U(1)$ coupling constants, which gives us an upper limit for the lightest Higgs particle which is just above $M_Z$ \cite{higgs}.  In the standard model, there is no such limit.  The parameters of the theory are the three gauge coupling constants and the parameters of the superpotential:
\begin{equation}
{W} = q{\lambda}_{u} \bar{u}H_u + q {\lambda}_{d} \bar{d}H_d +
l {\lambda}_{e}\bar{e}H_d + \mu H_u H_d
\end{equation}
where  $\lambda_u$, $\lambda_d$, and $\lambda_e$ are the $3 \times 3$ Yukawa coupling matrices (all flavor indices are suppressed) and $\mu$ is the Higgs mixing parameter.  We are assuming the neutrinos are massless and that R-parity is conserved.

Since no superpartners have been observed, supersymmetry must be broken.  Early attempts to break supersymmetry spontaneously and communicate the breaking through tree level yielded an unacceptable particle spectrum.  The first realistic models of supersymmetry included mass terms which broke supersymmetry explicitly but softly, so that the cancellation of quadratic divergences is maintained \cite{dg}.  The allowed mass terms are gaugino masses, scalar masses, and the bi-linear and trilinear terms \cite{gg}.

The Gaugino masses are: 
\begin{equation}
{\cal{L}}_{\rm m,gaugino} = M_3 \tilde{G}^{a} \tilde{G}^{a} + M_2 \tilde{W}^{i} \tilde{W}^{i} + M_1 \tilde{B} \tilde{B} + h.c.
\end{equation}
where $\tilde{G}$ is the gluino; $\tilde{W}$ is the wino; and $\tilde{B}$ is the bino.   Throughout the paper, we will use the convention that the tilde indicates the superpartner to the standard model particle.   The scalar masses are:
\begin{equation}
\label{eq:smass}
{\cal{L}}_{\rm m,scalar} = \sum_{A,i,j}{ m^{2}_{Aij} \tilde{A}_{i}^{*} \tilde{A}_j } + m^2_{Hu}  H_{u}^{*} H_{u} + m^2_{Hd}  H_{d}^{*} H_{d}
\end{equation}
where $\tilde{A}=\tilde{q}$, $\tilde{\bar{u}}$, $\tilde{\bar{d}}$, $\tilde{l}$, $\tilde{\bar{e}}$ labels the five species that constitute a family;  and i,j = 1,2,3 are U(3) flavor labels.  The bi-linear and tri-linear terms are:
\begin{equation}
{\cal{L}}_{\rm m,triscalar} = \tilde{q} {A'}_{u}  \tilde{\bar{u}}H_u + \tilde{q}  {A'}_{d} \tilde{\bar{d}}H_d +
\tilde{l} {A'}_{e} \tilde{\bar{e}}H_d + B' H_u H_d + h.c.
\end{equation}
where ${A'}_u$, ${A'}_d$, and ${A'}_e$ are  $3 \times 3$  matrices (flavor indices are suppressed) and $B'$ is a simple mass term\footnote{There are two conventions for bi-linear and tri-linear SUSY violating terms.  Here we define $A'$ and $B'$ to be the entire coupling.  The other, more common convention is that the entire coupling is obtained by multiplying the superpotential coupling, $\lambda$ or $\mu$, by the coefficients $A$ or $B$.  We will use both conventions in this paper, and they will be differentiated by the prime.  We use $A'$ and $B'$ here to emphasize the unconstrained nature of the terms.  After this section on the general SSM, we will use $A$ and $B$.}.

In the most general model, the supersymmetry breaking terms are arbitrary complex parameters subject only to the symmetries of the theory and experimental constraints.  We expect the scale of these masses to be near the weak scale based on naturalness.  There is no theoretical requirement of universality of the scalar masses and the $A$ terms do not have to be proportional to the Yukawa couplings.  We will refer to this general model as the supersymmetric standard model, or the SSM.

Although no supersymmetric partners have been observed, the success of the weak mixing angle prediction \cite{gutsin} in a SUSY GUT is strong support for low energy supersymmetry with unification at $M_{\rm GUT} \approx 10^{16}$ GeV.  In fact this is the only success of physics beyond the standard model.  Maintaining this prediction implies that there are no additional incomplete $SU(5)$ multiplets below the GUT scale\footnote{An $SU(5)$ multiplet does not effect the weak mixing angle prediction, but it does change the size of the coupling constant at unification.  The amount of new matter could be limited by requiring the coupling constants do not become strong between the weak and GUT scales.}.   

\subsection{Counting Parameters}
The gauge sector of the SSM contains three real gauge couplings and three complex gaugino masses.  We can do a continuous R-rotation to redefine one phase in the gaugino masses and the flavor sector.  We will use this freedom to make the Gluino mass real.  This gives us three dimensionless couplings, three masses, and two phases in the gauge sector.

The Higgs sector of the SSM contains the two SUSY violating scalar masses $m^2_{Hu}$ and $m^2_{Hd}$ which are real, and the complex parameters $\mu$ and $B$.  By doing a Peccei-Quinn phase redefinition, we can make one of these parameters real.  It is convenient to make $B$ real because this will make $\tan{\beta}$, the ratio of the Higgs VEVs, real.  There are thus four masses and one phase in the Higgs sector.

The remaining part of the Lagrangian, the flavor sector, contains three fermion Yukawa matrices, three triscalar coupling matrices, and five scalar mass matrices.  The Yukawa and triscalar  matrices are general $3 \times 3$ matrices with nine real magnitudes and nine imaginary phases each.  The five scalar mass matrices are $3 \times 3$ Hermitian matrices with six real magnitudes and three phases each.  This gives a total of 84 real parameters (mass eigenvalues and angles) and 69 phases.

Not all of these parameters are physical.  The gauge and Higgs sectors are invariant under a $U(3)$ flavor rotation for each of the five different types of particles:  $q$, $\bar{u}$, $\bar{d}$, $l$, and $\bar{e}$.  The parameters of the flavor sector violate this symmetry so it can be used to remove some of these parameters.  The $U(3)^5$ group has 15 angles and 30 phases; however, the flavor sector is invariant under two of the phases redefinitions, corresponding to baryon and lepton number.  Therefore, a total of 15 angles and 28 phases can be removed\footnote{In doing a similar counting for the standard model, the Lagrangian is invariant under four $U(1)$ field redefinitions:  baryon number and the three individual lepton numbers.}. 

Subtracting the removable parameters, we find the flavor sector contains 69 real parameters and 41 phases. Of the 69 real parameters 30 are masses, nine for fermions and 21 for scalars, and the remaining 39 are accounted for by a set of non-independent mixing angles and tri-linear couplings.  In the gauge and Higgs sector, we have the additional three coupling constants, seven masses, and three phases counted above, and the parameter $\theta_{\rm QCD}$.   

In this paper, we will be concerned with the new flavor parameters, and in particular, the real parameters.  Compared to the standard model, there are an additional 21 masses, 36 mixing angles and 40 phases in the flavor sector. They all imply new physics. A geometric interpretation of these parameters will be given in the next section.

\subsection{Sparticle Basis}
The soft scalar masses, the first term of equation~\ref{eq:smass}, are quadratic and chirality conserving.  A $U(3)^5$ rotation can diagonalize it and take us to the ``sparticle'' basis \footnote{Unless otherwise specified we will always make superfield rotations: sparticles and particles are rotated in parallel. This ensures that the gaugino couplings have their minimal form.} where:
\begin{equation}
 m^{2}_{Aij} =  m^{2}_{A(i)} \delta_{ij}
\end{equation}

Thus, in this basis, these chirality conserving terms of the Lagrangian also conserve a $U(1)^{15}$  flavor subgroup that conserves individual species number for each of the 15 species of quarks and leptons that make up the three families. In the sparticle basis, although the chirality conserving terms in the Lagrangian distinguish the 15 species of sparticles, they do not cause flavor violating transitions between them. This is convenient for tracing flavor violations; they are associated with chirality violations and originate either in the Yukawa superpotential or in the triscalar couplings.

In this basis the Yukawa superpotential has the form:
\begin{equation}
{W}_{\rm Yukawa} = q U_{q} \bar{\lambda}_{u} U_{\bar{u}} \bar{u} H_{u} + q U_{q}' \bar{\lambda}_{d} U_{\bar{d}} \bar{d} H_{d} +
l U_{l} \bar{\lambda}_{e} U_{\bar{e}} \bar{e} H_{d}
\end{equation}
\noindent where $\bar{\lambda}_{u}$,$\bar{\lambda}_{d}$, and $\bar{\lambda}_{e}$ are the diagonal Yukawa couplings for the quarks and electrons.  $U_{q}'$, $U_{q}$, $U_{\bar{u}}$, $U_{\bar{d}}$, $U_{l}$, and $U_{\bar{e}}$ are six unitary matrices; $U_{q}^{\dagger} U_{q}'$ is the usual KM matrix, whereas the remaining five are new independent matrices.  In general, these matrices cannot be rotated away. They have both physical and geometrical significance. Their physical significance is that they cause new flavor violations. Their geometrical significance is that they measure the relative misalignment between sparticle and particle masses in flavor $U(3)^5$ space.  The A-terms are of a similiar form to the Yukawa couplings.  They contain six additional $3 \times 3$ unitary matrices with a similiar physical interpretation.

\subsection{Universality and Proportionality}
In minimal supersymmetric theories it is often assumed that, at some fundamental scale  $\sim M_{\rm GUT}$ or $M_{\rm string}$, each triscalar coupling is proportional to the corresponding Yukawa coupling with a proportionality constant which is the same for each Yukawa matrix. This is sometimes called {\it proportionality} and reduces the possible 27 complex numbers to one.  In addition, again in minimal theories, one of two conditions is also postulated \cite{dg}:
\begin{itemize}
\item {{\bf horizontal universality}:
\begin{equation}
 m^{2}_{Aij} =  m^{2}_{A} \delta_{ij}
\end{equation}}
or, the more restrictive
\item {{\bf universality}:
\begin{equation}
 m^{2}_{Aij} =  m^{2} \delta_{ij}
\end{equation}}
\end{itemize}

Either version of universality reduces the sparticle masses to spheres in flavor space which preserve the full $U(3)^5$ rotation group. Since a sphere points nowhere, the notion of relative orientation of particle and sparticle masses loses its meaning; the geometric significance of 5 of the 6 matrices $U_{q}'$, $U_{q}$, $U_{\bar{u}}$, $U_{\bar{d}}$, $U_{l}$, and $U_{\bar{e}}$   disappears. Only the usual CKM matrix $U_{q}^{\dagger} U_{q}'$ that measures the relative orientation of up and down quark masses continues to have geometrical and physical meaning. In particular, since $U_{l}$ and $U_{\bar{e}}$ lose their meaning, there are no lepton number violations in theories satisfying horizontal universality.
The importance of the hypotheses of proportionality and universality is now clear: They insure that all flavor violations involve the quarks and are proportional to the usual CKM matrix $U_{q}^{\dagger} U_{q}'$; consequently, they are under control.

As we shall review in the next section, the problem with these hypotheses is that they do not seem to emerge from fundamental short distance theories, such as GUTs or strings: Flavor breakings in the fermion sector invariably pollute the soft terms and render them non-universal and non-proportional.  This is, in one sense, fortunate because these low energy parameters may serve as a fingerprint of high energy physics that is otherwise beyond the reach of experiment. 

Since we wish to do a general analysis of flavor violations we will not assume proportionality or any form of universality.  We will however assume all the parameters are real.

\section{Sources of Non-Universality}
\label{sec:nonun}

All theories have some degree of flavor dependence in the soft SUSY breaking terms.  The terms which violate the $U(3)$ flavor symmetries for the fermions will also affect the soft terms.  The key question is the extent to which the sparticles are non-degenerate between families and misaligned with respect to the fermions.  In this section we will consider flavor dependence in the soft terms induced from the superpotential through loop effects \cite{hkr}\footnote{Flavor dependence may arise through other methods.  For example, a broken flavor symmetry can cause nonuniversality in the scalar masses \cite{dterm} or the scalar masses may be originally generated with flavor dependence, which is the generic case of supergravity breaking \cite{kl}.}.  The important factors are the superpotential couplings and the nature of the supersymmetry breaking.

In the most common form of supersymmetry breaking used in model building, supersymmetry is broken in a hidden sector that contains no couplings to the standard model particles except through gravity \cite{sugra}.  The resulting soft terms can be flavor blind at $M_{\rm Planck}$; however, we must use the RG equations to relate the $M_{\rm Planck}$ values to the $M_{\rm weak}$ values used in calculations.  The Yukawa couplings appearing in the RG equations between $M_{\rm GUT}$ and $M_{\rm weak}$ do not cause significant problems.  The unknown Yukawa couplings between $M_{\rm Planck}$ and $M_{\rm GUT}$ however can be dangerous.

In theories that do not explain the flavor hierarchy, most of the Yukawa couplings are small so they do not contribute significantly to flavor violations.  The top Yukawa, however, is large, so it can induce measurable violations \cite{bh}.  Recent calculations of $\mu \rightarrow e + \gamma$ in the minimal SUSY GUT give results that could be observed soon if sparticles are not too heavy \cite{bh}.  In this case, the large top Yukawa does not cause a bigger problem because it is sheltered from the first generation by an additional small mixing angle and because it only creates a mass splitting among the right handed electrons, which does not give the major contribution to $\mu \rightarrow e + \gamma$.

In theories that do explain the flavor hierarchy \cite{an}, there are no small parameters and that  all non-vanishing Yukawa couplings are of the same order as the gauge coupling at some high scale $ \sim  M_{\rm PL}$ or $ M_{\rm GUT}$, which we call the flavor scale.  These Yukawas couple the three ordinary families to superheavy multiplets residing at the flavor scale.  As we shall demonstrate, they can create large splittings among the ordinary squarks and sleptons which subsequently lead to dangerous flavor violating interactions.  Even if there is a flavor symmetry protecting the soft terms, threshold corrections will occur when the symmetry is spontaneously broken, resulting once again in dangerous contributions.

Let us examine the effects of a single large coupling that is asymmetric between the families.  We will look at a unified theory in which all the low energy families have identical couplings except one family has an additional coupling to two particles with masses at the GUT scale, as shown in equation~\ref{eq:sup}.
\begin{equation}
\label{eq:sup}
W_{\rm assym} = \lambda \bar{5}_{\rm light} \bar{5}_{\rm GUT} 10_{\rm GUT}
\end{equation}

In the linear approximation, we can find the induced mass splitting by using the RG equation obtained from considering only the asymmetric term.
\begin{equation}
\label{eq:ev}
\frac{dm_{\rm family}^2}{dt} = \frac{1}{8 \pi^2} 4 \lambda^2 ( m_{\rm family}^2 + m_{\rm \bar{5}heavy}^2 + m_{\rm 10heavy}^2 + A^2)
\end{equation}
We will assume all the SUSY breaking scalar masses have a common value, and that the $A$ parameter has that same value.  We will also set $\lambda=1$.  We will assume this equation is valid starting at the string scale, $5 \times 10^{17}$ GeV, until the GUT scale, $2 \times 10^{16}$, where we integrate out the GUT particles.  The change in the mass obtained, which is equal to the mass splitting, is $\delta m^2= .65 m^2$ in the linear approximation.  Clearly the linear approximation breaks down, but we do expect fractional splittings of O(100\%) if there are large Yukawa couplings over a broad range of energies.

Let us redo the calculation, now assuming the heavy particles in the asymmetric coupling have a mass $1/2 M_{\rm string}$ instead of $M_{\rm GUT} = 1/25 M_{\rm string}$ as above.  Integrating the RG equation over the range  $M_{\rm string} \rightarrow 1/2 M_{\rm string}$ is a typical approximation to a threshold correction from a broken symmetry, in this case a flavor symmetry.  The induced mass splitting is $\delta m^2=.14 m^2$.  This is still large, as we shall see from flavor changing calculations.  Because there is a logarithmic dependence on the ratio of mass scales, even a small integration interval gives a significant mass correction.

The above examples of induced flavor dependence in the soft terms have assumed supersymmetry breaking is communicated to the standard model particles by Planck mass particles.  If this communication occurs through GUT mass particles, then the RG evolution from the string scale is not valid.  However, we can still have a threshold effect contribution at the GUT scale.  The unsuccessful GUT predictions $e/\mu = d/s$ and $\mu/\tau = s/b$ are evidence of non-trivial flavor physics at the GUT scale, which could be the source of threshold corrections. 

In any theory of soft SUSY breaking terms there must be a violation of the flavor symmetry communicated through loop effects due to the flavor breaking Yukawa sector.  Small Yukawa couplings are not dangerous, but in the presence of large Yukawa couplings, as expected near the flavor scale, there will be large flavor violations in the SUSY breaking sector.

\section{Naturalness}

The naturalness criterion measures the sensitivity of the weak scale to variations of the SUSY parameters at a fundamental scale \cite{bg}\cite{fine2}.  Here we will use the GUT scale since we are assuming that is the scale of new physics.  In this section we will review a simplified form of the analysis of Barbieri and Giudice \cite{bg}.

If the conditions for symmetry breaking are met, the minimum of the Higgs potential at tree level can be written in terms of two equations:
\begin{equation}
\sin{2 \beta} =  \frac{2 B \mu }{(m_{Hd}^2 + \mu^2) + (m_{Hu}^2 + \mu^2)}
\end{equation}
\begin{equation}
\label{eq:mz}
M_z^2 = 2 \frac{ (m_{Hd}^2 + \mu^2) -  (m_{Hu}^2 + \mu^2) \tan{^2 \beta} }{ \tan{^2 \beta} - 1 }
\end{equation}
\noindent Here $m_{Hd}^2$ and $m_{Hu}^2$ are the soft scalar masses of the down and up Higgs respectively, $\mu$ is the Higgsino mass, and $B \mu$ is the coupling from the SUSY violating term $B \mu H_u H_d$.  All parameters in the above equation are evaluated at $M_z$.

The next step is to write the equation for $M_Z$, equation~\ref{eq:mz}, in terms of parameters at the GUT scale, for which one loop RG equations are sufficient.  Here we will make a simplification from Barbieri and Giudice.  We will keep $\tan{\beta}$, evaluated at the weak scale, in the equation as a fundamental parameter.  This simplifies the resulting equation, making $M_z^2$ linear in the GUT scale parameters, which allows for an easier interpretation of the results.  The relevant numerical results are unchanged.  In addition, we will keep the $\mu$ parameter evaluated at $M_{\rm weak}$.  This does not effect the results because $\mu$ is renormalized only by a multiplicative constant.

Equation~\ref{eq:mz2} gives $M_z$ in terms of the parameters of interest.
\begin{eqnarray}
\label{eq:mz2}
M_z^2 & = & c_{\mu} \mu^2 + c_{Hd} m_{Hd0}^{2} + c_{Hu} m_{Hu0}^{2} + c_{t}   m_{t0}^{2} + c_{\bar{t}} m_{\bar{t}0}^{2} +
\nonumber \\
& & c_{M} M_0^2 + c_{AM} A_{t0} M_0 + c_{A} A_{t0}^{2}
\end{eqnarray}  
\noindent A subscript {\it 0} refers to a parameter evaluated at the GUT scale\footnote{We will keep this convention throughout the paper.}.  $M$ is the gaugino mass (unification is assumed), and $m^2$ is a soft scalar mass.  The $c$ coefficients are functions of $\tan{\beta}$ and constants of O(1) from RG solutions.  We have assumed the top Yukawa is the only contributing Yukawa coupling.

There is no {\it a priori} relation among the $c$ coefficients, so it is unlikely that a large cancellation between seperate terms of equation~\ref{eq:mz2} will occur.  We define the fine tuning of a given term as the fraction by which $M_z^2$ is smaller than that term.  For example, the fine tuning of the term associated with the parameter $\mu$, which we label $f_{\mu}$, is given in equation~\ref{eq:ftmu}.
\begin{equation}
\label{eq:ftmu}
f_{\mu} = \frac{M_z^2}{c_{\mu} \mu^2}
\end{equation}
\noindent  Unless there is some cancellation mechanism, the limit to a reasonable cancellation is usually placed at a fine tuning of $f=.1$.  This means the apparently unrelated terms of the equations for $M_z^2$, equation~\ref{eq:mz2}, conspire to sum to a value one tenth the value of the individual terms.  This is the 10\% naturalness criterion.

This analysis gives especially tight constraints on the parameters $\mu$ and $M_0$.  Independent of $\tan{\beta}$ and the renormalization group, the coefficient $c_{\mu}=2$.  The minimum value of $c_{M}$ is $\approx 6$ and occurs for $\tan{\beta} \gg 1$.  There is only a weak dependence here on the exact size of the top Yukawa because we are near the fixed point.  With these $c$ values, the 10\% fine tuning criterion gives the following upper mass limits:
\begin{eqnarray}
\nonumber
M_0 = 117 GeV
\\
\nonumber
\mu = 203 GeV
\end{eqnarray}
The constraints for the other particles are less restrictive.  Although we are near the top fixed point, the coefficient of the scalar masses $m_0^2$ passes through zero for a particular value of the top Yukawa in the experimental range.  At this (fine tuned) value of $\lambda_t$, $M_z^2$ is independent of $m_0^2$.

If one wants to allow for a larger fine tuning, the square of the masses can be scaled up by the factor by which the fine tuning is increased.  For example, a $1\%$ fine tuning gives an upper limit on $M_0$ of 370~GeV.

\section{Flavor Changing Processes}
\label{sec:fc}

We will now calculate the constraints on flavor dependence in the soft supersymmetry breaking terms by examining the flavor changing processes $\mu \rightarrow e + \gamma$ and $K^0 - \bar{K}^0$ mixing.  To do this, we can make two simplifying approximations which we will justify below;  we will neglect the A terms and the third family.

In order to protect $\mu \rightarrow e + \gamma$ from a large $A$ term contribution, we must assume the $A$ terms are {\it approximately} proportional, just as we must also assume the scalar masses are approximately universal.  Given approximate proportionality, with splittings among the $A$ entries comparable to the splittings among the scalar mass terms, individual $\mu \rightarrow e + \gamma$  diagrams involving $A$ will be equal in size to the leading diagrams.  However, there are several leading diagrams.  When all the diagrams are added together, the $A$ diagrams make a qualitatively unimportant contribution, justifying the approximation $A=0$.  In $K^0 - \bar{K}^0$ mixing, approximate proportionality renders the $A$ contribution unimportant.

We can neglect the third family contribution because we are interested in the flavor violation resulting from the flavor scale.  The third family contribution is suppressed by a small, extra mixing angle so it will only be important if the mass splittings involving the third family are much larger than the mass splitting between the first two families.  On the contrary, the mass splittings resulting from the flavor scale should be of the same size for all three families
\begin{equation}
m^2_{1} - m^2_{2} \approx m^2_{2} - m^2_{3} \approx m^2_{3}-m^2_{1}
\end{equation}

\noindent so the relevant third family contribution will be unimportant.

In the low energy theory, we do expect an additional contribution to the non-universality in the third generation coming from the large third family Yukawa below the flavor scale.  However, since we are trying to put an upper bound on the non-universality originating at the flavor scale, we will assume all the flavor violation originates there.  Any additional contribution will only strengthen the bounds on the physics of flavor.

\subsection{$\mu \rightarrow e + \gamma$}
\label{sec:mass}

We will use the $\mu \rightarrow e + \gamma$ branching ratio calculation of chapter~\ref{chap:calc}, which calculates all leading one loop contributions.  Previous analyses omitted several significant contributions, often even the largest ones.

Neglecting the $A$ terms and the third family, the calculation includes only three $2 \times 2$ mass matrices: the Yukawa matrix, the lepton doublet scalar mass matrix, and the electron singlet scalar mass matrix.  The associated physical parameters are the two Yukawa eigenvalues; two scalar mass eigenvalues for both the lepton doublet and electron singlet; and a mixing angle for both the lepton doublet, $\theta_{l}$, and electron singlet, $\theta_{\bar{e}}$, that describes the rotation between the sparticle and particle mass eigenbases.

Because the scalar mass splittings are required to be small, we will parametrize the doublet and singlet scalar mass eigenvalues by the average masses, $m_{l}^2$ and $m_{\bar{e}}^2$, and the mass splittings, $\delta \tilde{m}_{l}^2$ and $\delta \tilde{m}_{\bar{e}}^2$.  We will also keep only the leading contribution in both the mass splittings and the mixing angles.
Equation~\ref{eq:gamp} gives the branching ratio for the process $\mu \rightarrow e+\gamma$.  The functions $X_{l}$ and $X_{\bar{e}}$ are given in appendix~\ref{sec:functions}.
\begin{equation}
\label{eq:mue}
BR(\mu \rightarrow e+\gamma)= \frac{3e^{2}}{2\pi^{2}} \left\{ \theta_{l}^{2} \left( \frac{M_{w}}{m_{\tilde{l}}} \right)^{4} (X_{l}) ^{2} \left( \frac{\delta \tilde{m}_{l}^2}{m_{\tilde{l}}^{2}} \right)^2 + \theta_{\bar{e}}^{2} \left( \frac{M_{w}}{m_{\tilde{\bar{e}}}} \right)^{4} (X_{\bar{e}}) ^{2} \left( \frac{\delta m_{\tilde{\bar{e}}}^2}{m_{\tilde{\bar{e}}}^{2}} \right)^2 \right\}
\label{eq:gamp}
\end{equation}

\subsection{$K^{0}-\bar{K}^{0}$}

We will use the $K^{0}-\bar{K}^{0}$ mixing calculation of reference \cite{hag} which computed the dominant supersymmetric contribution, the gluino box diagrams.  Because there are no charged currents, the weak singlet up quark does not appear in our calculation.  The parameters in this calculation are the same as in the $\mu \rightarrow e + \gamma$ calculation with the quark doublet and down quark singlet replacing the lepton doublet and electron singlet.  We will therefore use a parallel notation.  The $q$ subscript refers to the quark doublet, and $\bar{d}$ refers to the down quark singlet.

The kaon mass splitting is given in equation~\ref{eq:kkbar}.  The definitions of $f_{1}$ and $f_{2}$ are given in appendix~\ref{sec:functions}.
\begin{eqnarray}
\label{eq:kk}
\Delta M_{K}= \frac{\alpha_{s}^{2}}{216 m_{\tilde{q}}^{2}} \left( \frac{2}{3} f_{K}^{2} m_{K} \right) \left\{ \theta_{dl}^{2} \left( \frac{\delta \tilde{m}_{\tilde{q}}^2}{m_{\tilde{q}}^2} \right)^2 f_{1} \! \left( \frac{M_{\tilde{g}}^2}{m_{\tilde{q}}^2} \right) + \right. \nonumber \\ \left. \theta_{q}\theta_{\bar{d}} \left( \frac{\delta m_{\tilde{q}}^2}{m_{\tilde{q}}^2} \right) \left( \frac{\delta \tilde{m}_{\bar{d}}^2}{m_{\tilde{\bar{d}}}^2} \right) f_{2} \! \left( \frac{M_{\tilde{g}}^2}{m_{\tilde{q}}^2} \right) + \theta_{\bar{d}}^{2} \left( \frac{\delta \tilde{m}_{\bar{d}}^2}{m_{\tilde{\bar{d}}}^2} \right)^2 f_{1} \! \left( \frac{M_{\tilde{g}}^2}{m_{\tilde{\bar{d}}}^2} \right) \right\}
\label{eq:kkbar}
\end{eqnarray}

\subsection{Experimental Constraints}

\begin{figure}
	\epsfysize 16cm
	\centerline{\epsffile{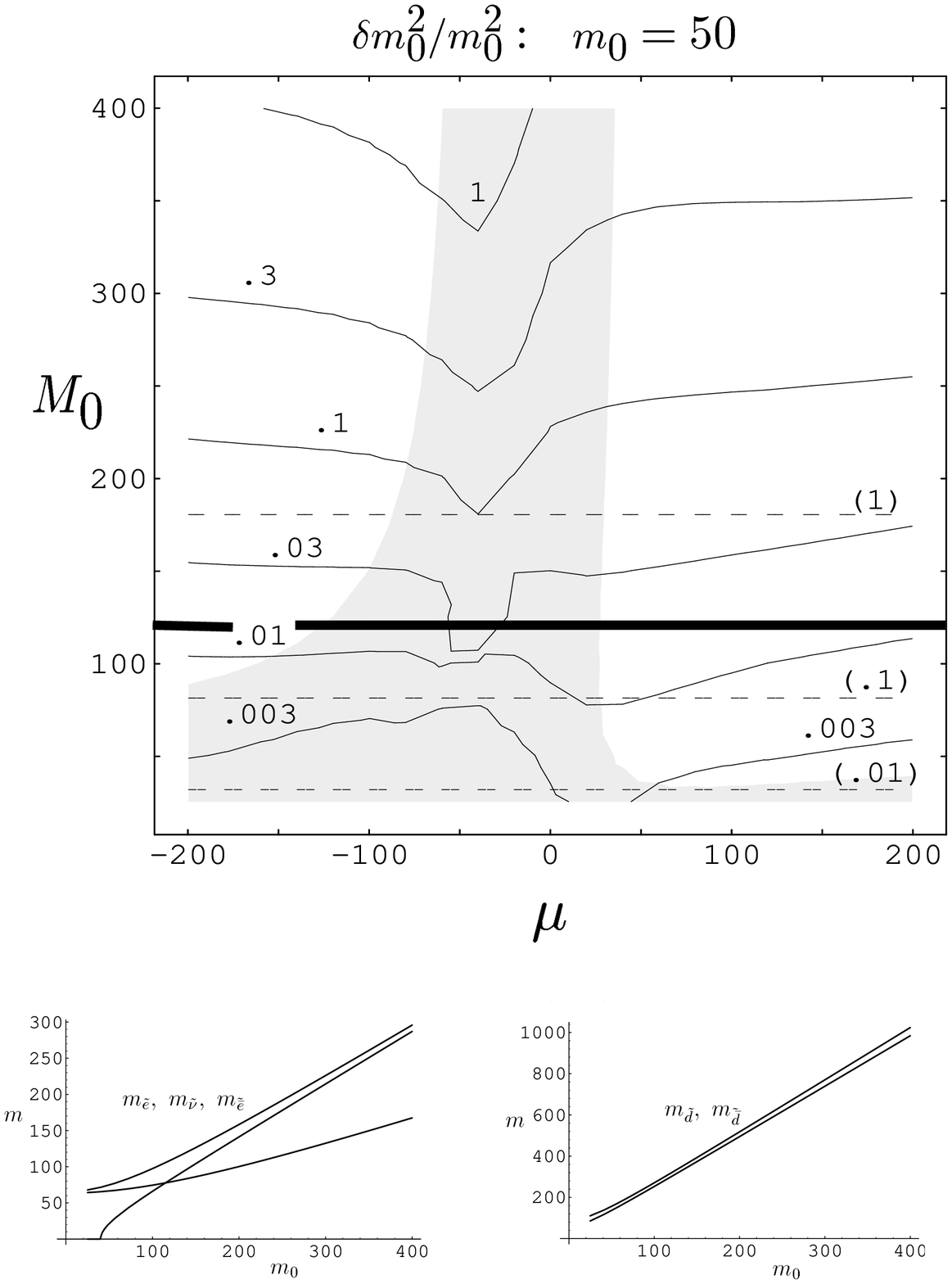}}
	\caption{Fractional mass splittings constraints for $m_0=50$ GeV
	(top).  Physical masses of sleptons (bottom left) and down squarks
	(bottom right).}
	\label{fig:m50}
\end{figure}

\begin{figure}
	\epsfysize 16cm
	\centerline{\epsffile{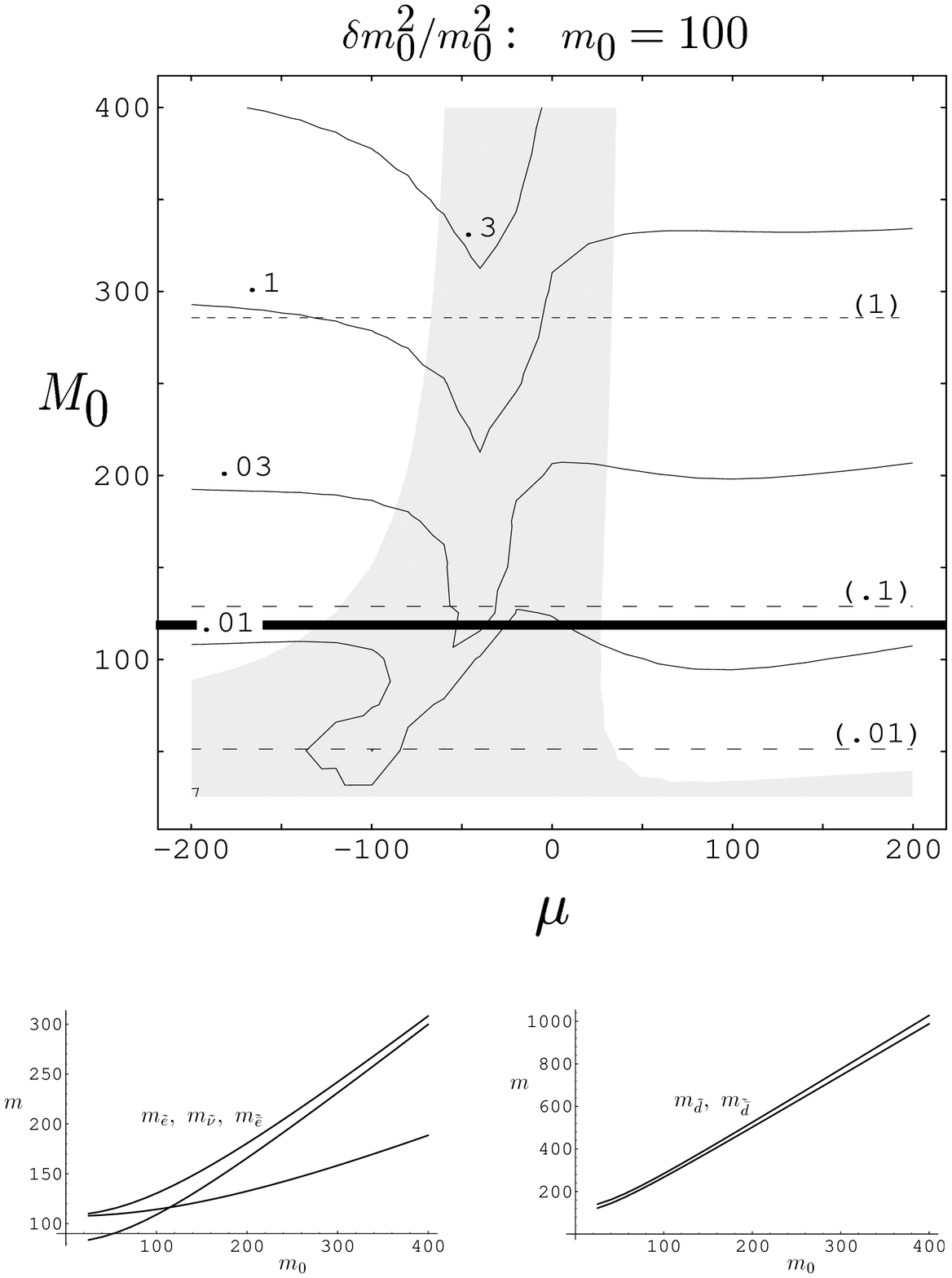}}
	\caption{Fractional mass splittings constraints for $m_0=100$ GeV
	(top).  Physical masses of sleptons (bottom left) and down squarks
	(bottom right).}
	\label{fig:m100}
\end{figure}

\begin{figure}
	\epsfysize 16cm
	\centerline{\epsffile{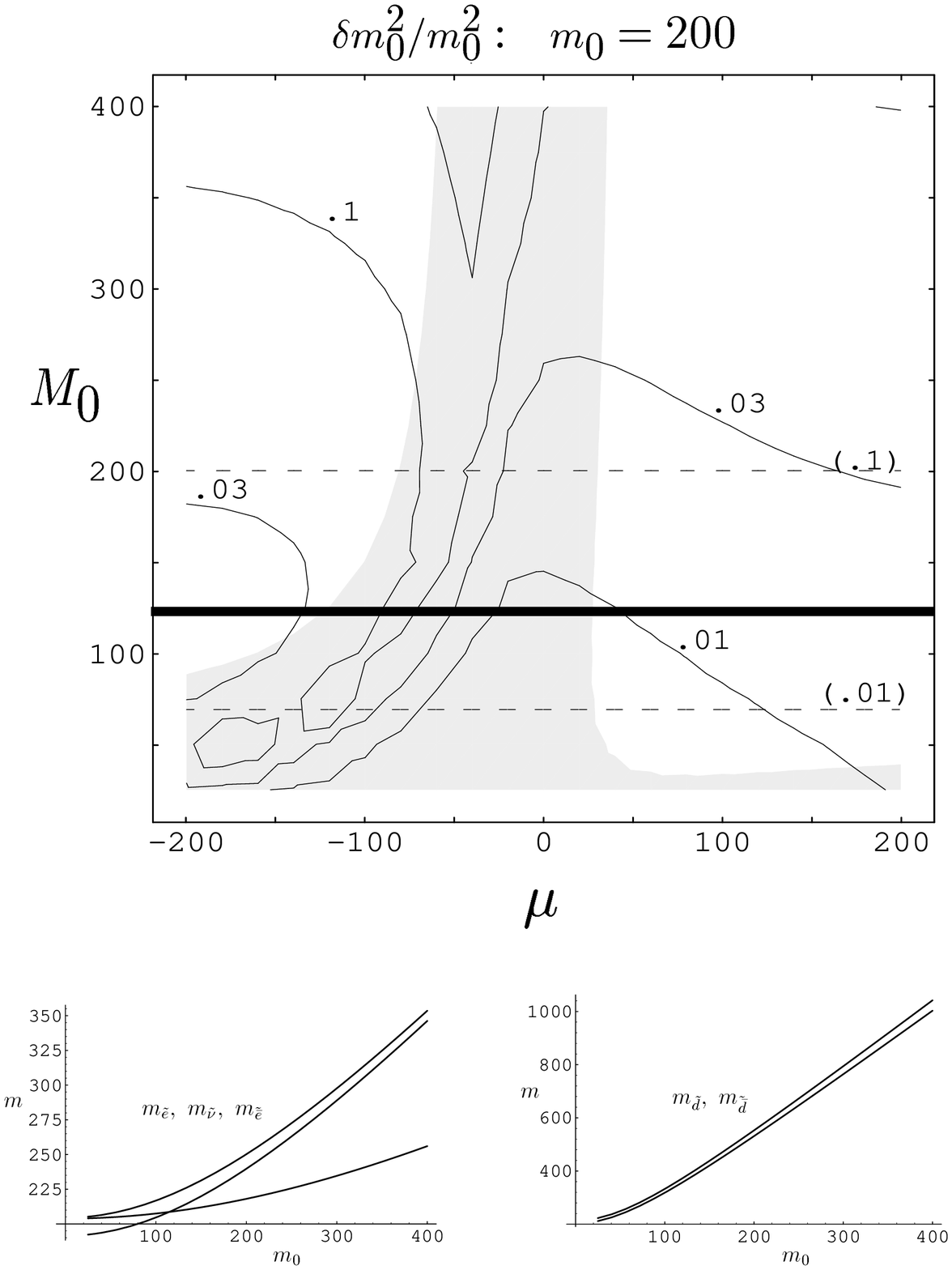}}
	\caption{Fractional mass splittings constraints for $m_0=200$ GeV
	(top).  Physical masses of sleptons (bottom left) and down squarks
	(bottom right).}
	\label{fig:m200}
\end{figure}

\begin{figure}
	\epsfysize 16cm
	\centerline{\epsffile{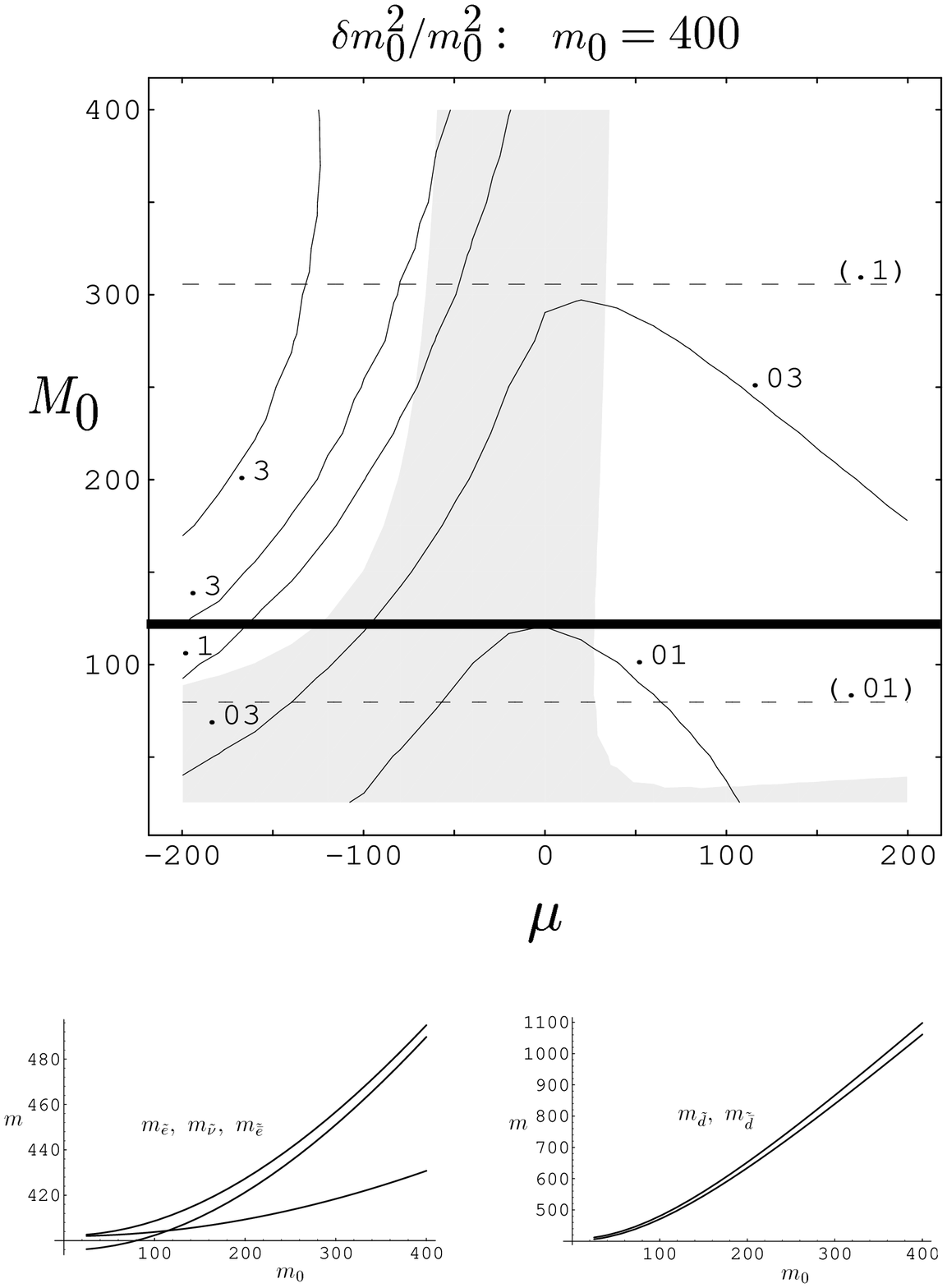}}
	\caption{Fractional mass splittings constraints for $m_0=400$ GeV
	(top).  Physical masses of sleptons (bottom left) and down squarks
	(bottom right).}
	\label{fig:m400}
\end{figure}

We will take equations~\ref{eq:mue}~and~\ref{eq:kk} and solve them for the fractional scalar mass splitting $\delta m^2 / m^2$.  We then use the one loop RG equations \cite{rg} to relate the low energy result to the fundamental scale, which we assume is $M_{\rm gut}$ for the graphs of figures~\ref{fig:m50}-\ref{fig:m400}.  Because we ignore the contribution of the two lightest generation Yukawa couplings, the only source of mass splitting between the first and second family will be the boundary conditions at the GUT scale.

For the gauge coupling constants, we take as inputs $\sin{^{2}\theta_{w}}=.232$ and $\alpha_{em}=1/127.9$.  We assume the coupling constants unify at the GUT scale using the one loop equations, but we use a larger value of $\alpha_{s}=.12$ for the sake of low energy calculations.  The experimental inputs are $BR(\mu \rightarrow e+\gamma)=4.9 \times 10^{-11}$ and $\Delta M_{K} / M_{K}=.71 \times 10^{-14}$. \cite{pdb}   

For our graphs we choose $\tan{\beta}=3$ and, as stated above, $A=0$.  The mass splitting constraint gets stricter for larger values of $\tan{\beta}$, but does not change much as $\tan{\beta}$ gets smaller.  In appendix~\ref{app:Atan}, we show graphs displaying the $\tan{\beta}$ and $A$ dependence.

For simplicity of presentation, we assume the singlet and doublet mixing angles and mass splittings are the same.  Furthermore, we assume each mixing angle is equal to the square root of the masses of the two particles it relates: for the leptons, $\theta_{l}=\theta_{\bar{e}}=\sqrt{e/\mu}$, and for the quarks, $\theta_{q}=\theta_{\bar{d}}=\sqrt{d/s}$.  The result holds, at least approximately, in most unified theories of fermion masses \cite{an} and is a consequence of  quark-lepton unification and the successful relation: $\theta_{\rm Cabibbo}=\sqrt{d/s}$.

Figures~\ref{fig:m50} through~\ref{fig:m400} are contour plots of the upper limits on the fractional scalar mass splittings, evaluated at the GUT scale, as a function of SUSY parameter space.  We show four graphs, one for each of four values of the scalar mass evaluated at the GUT scale, $m_0$.  The axes of the graphs are the Higgs mixing parameter evaluated at the weak scale, $\mu$, and the gaugino mass evaluated at the GUT scale, $M_0$. The solid contours are the upper limit of the fractional mass splitting of the sleptons from $\mu \rightarrow e+\gamma$.  The dashed lines, which are labeled in parentheses, are the upper limit of the fractional mass splitting of the down and strange squarks from $K^{0}-\bar{K}^{0}$ mixing.  We have also included a bold line at $M_0=120$ GeV which is the maximum value of $M_0$ based on the 10\% naturalness criterion \cite{bg}.  The shaded region is the experimentally excluded region where the lightest chargino is less than 45 GeV. 

Accompanying each contour plot, we have included two graphs which give the associated physical masses of the three sleptons and two down squarks as a function of $M_0$.

These constraints can easily be adapted for new values of the mixing angles or the branching ratio.  The resulting fractional mass splittings may be read from the contour graphs using equations~\ref{eq:out1} and~\ref{eq:out2}.
\begin{eqnarray}
\label{eq:out1}
\left( \frac{\delta m_0^2}{m_0^2} \right)_{\rm lepton} & = (\frac{\delta m_0^2}{m_0^2}~from~graph)_{\mu \rightarrow e+\gamma} \left\{ \frac{BR(\mu \rightarrow e+\gamma)}{4.9 \times 10^{-11}} \right\}^{\frac{1}{2}} \left\{ \frac{.07}{\theta_{\rm lepton}} \right\}
\\
\label{eq:out2}
\left( \frac{\delta m_0^2}{m_0^2} \right)_{\rm down} & = (\frac{\delta m_0^2}{m_0^2}~from~graph)_{K^{0} - \bar{K}^{0}} \left\{ \frac{ \Delta M_{K} / M_{K} }{ .71 \times 10^{-14} }  \right\}^{\frac{1}{2}}\left\{ \frac{.22}{\theta_{\rm down}} \right\}
\end{eqnarray}

If we take the upper limit of $M_0=120$ GeV from the 10\% fine tuning criterion, we see that the upper limit to the slepton fractional mass splitting is about .01 on all the graphs.  There is an exception to this for large values of $m_0$ (400 GeV) and a negative value for $\mu$.  The amplitude for the decay to a final state left handed electron passes through zero here, leaving only the less important right handed electron contribution and making the limit not as strong.  However, the $K^0-\bar{K}^0$ mixing constraint is important in this range of parameter space, and we still obtain a mass splitting limit near .01, this time for the down-type quarks.  In light of the mass splitting induced by a threshold correction at the flavor scale (section~\ref{sec:nonun}), this mass splitting is unnaturally small.

We can obtain more reasonable mass splitting limits if we relax the $M_0=120$ GeV constraint.  If we allow $M_0 \approx 300$ GeV or $400$ GeV, the fractional mass splitting constraints are weakened to $\approx$ .1 to .3.  However, this requires fine tuning of 1\% from the naturalness criterion.  In other words, the apparently unrelated terms in the equation for electroweak breaking, equation~\ref{eq:mz2}, sum to give an answer 100 times smaller than the individual terms.  This is difficult to swallow unless a cancellation mechanism exists.

We can not simultaneously satisfy constraints from naturalness and flavor differentiation.  This implies that there must be a mechanism that suppresses the supersymmetric contribution to flavor changing processes.

\section{A Case for Light Messengers}

The present paper has focused on the conflict between the following statements:
\begin{itemize}
\item[1)] Theories attempting to explain (even small parts of) the flavor hierarchy predict a large flavor dependence in the sparticle mass matrices;
\item[2)] Naturalness implies light sparticles;
\item[3)] Suppression of rare processes implies that either sparticle mass splitting and mixing angles are small, or sparticle masses are large.
\end{itemize}

One way to avoid this conflict is to have a theory of flavor that does not generate large mass splittings or mixing angles, evading statement (1).  Examples in the literature addressing this issue are the dynamical alignment of the sparticle and particle mass matrices \cite{dgiu} and the non-abelian flavor symmetry \cite{ps}.  Another way to avoid the conflict is to have the breaking of supersymmetry such that the SUSY violating terms are decoupled from the high energy physics, once again evading statement (1).

Consider, for example, a theory in which the soft terms shut off above a scale $\Lambda \ll M_{\rm PL}$ (or $M_{\rm GUT}$).  In such a theory the soft terms would not be distorted by the flavor physics that takes place at $\sim  M_{\rm PL}$  (or $M_{\rm GUT}$) and gives rise to the ordinary quark and lepton masses.  If the soft terms are generated at the scale $\Lambda \ll M_{\rm PL}$  and satisfy universality and proportionality then they will not cause any large flavor violations near the weak scale.  The deviations from universality and proportionality that arise between the scales $\Lambda$ and $ M_W$ are caused by the ordinary Yukawa couplings and are harmless.

\begin{figure}
\centerline{\epsffile{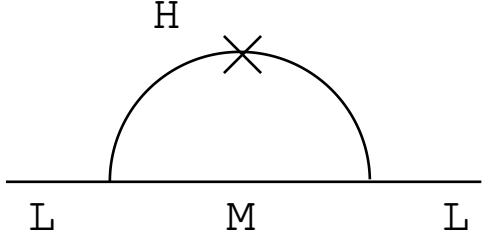}}
\caption{Schematic diagram illustrating how SUSY breaking is communicated from the hidden sector $H$ via a messenger $M$ to $SU_3 \times SU_2 \times U_1$ carrying sparticles $L$.  $L$ can be light $\sim M_W$ or heavy $\sim M_{\rm GUT}$.}
\label{fig:hidden}
\end{figure}

An interesting class of such theories are those with dynamically broken supersymmetry near the weak scale~\cite{dine}.  Another class are (scaled down versions of) the geometric hierarchy type theories \cite{dr}.  These are theories in which SUSY breaking originates in a hidden sector $(H)$ and is communicated to the particles carrying $SU_3 \times SU_2 \times U_1$ quantum numbers $(L)$ via messengers $(M)$ as pictured in Fig.~\ref{fig:hidden}.  The particles $L$ carry $SU_3 \times SU_2 \times U_1$ quantum numbers and can be light $\sim M_{\rm weak}$ or heavy $\sim M_{\rm GUT}$.

The soft masses induced by Fig.~\ref{fig:hidden} are, for example, of the form
\begin{equation}
\tilde m_L \simeq \frac{M^2_H}{M_M} \sim M_{\rm weak}
\label{soft}
\end{equation}
where $M_M$ is the messenger mass and $M_H$ is a SUSY breaking mass.  In the geometric hierarchy models \cite{dr}, $M_M \sim M_{\rm GUT}$ and $M_H \sim 3 \times 10^9$~GeV.  In models where the messenger is gravity \cite{sugra}  $M_M \sim M_{\rm PL}$ and $M_H \sim 3 \times 10^{10}$~GeV. It is not difficult to consider geometric hierarchy type models where the messenger mass $M_M$ is lighter than $M_{\rm GUT}$ and $M_H \sim \sqrt{M_M  M_{\rm weak}}$ is proportionally lighter.  What does one gain by this? At high momenta $p$ the soft term $\tilde m_L$ of the above equation behaves as:
\begin{equation}
\tilde m_L(p) \simeq \frac{M^2_H}{p}
\end{equation}

It shuts off at $p \gg  M_M$, and does not feel any of the flavor physics happening near $ M_{\rm PL}$ (or $M_{\rm GUT}$). Consequently, the sparticle splittings and rare processes coming from Planckian (or GUT) physics are suppressed by powers of $M_M/M_{\rm PL}$ (or $M_M/M_{\rm GUT}$) relative to their values in models where the messenger is supergravity. The phenomenology of such models is quite different from the canonical supersymmetric theories where $M_M \sim M_{\rm GUT}$ or $M_{\rm PL}$.  In particular, if $M_M\ll M_{\rm GUT}$, the sparticle masses are more degenerate and deviations from universality or proportionality are smaller. 

\section{Conclusion}

The minimal supersymmetric standard model with universality and proportionality provides a means of  preserving the light weak scale and it is consistent with the observed flavor changing data.  However, it does not seem to arise from a fundamental theory explaining flavor.  Models which do explain the fermionic flavor hierarchy generically contain flavor dependence in the sparticle masses which contributes to flavor changing interactions.  One solution to this problem is to increase the mass of the sparticles.  However, the sparticle masses have an upper limit which comes from naturalness of the electroweak breaking parameters.

In this chapter, we have reviewed the origin of flavor dependence in the scalar mass matrices and the fine tuning constraints (naturalness criterion) of electroweak breaking.  We then graphed the constraints from flavor changing processes for a general class of theories to quantify the conflict that exists between the physics of flavor and naturalness in electroweak breaking.  We have found that it is not possible to simultaneously satisfy the constraints from both flavor and naturalness, implying that there must be a new mechanism which allows the observed smallness in flavor changing processes.

There are two main approaches to controlling the supersymmetric contribution to flavor changing neutral currents by limiting the flavor dependence in the soft terms.  One approach is to have a theory of flavor which does not pollute the SUSY breaking terms, and the other is to have a theory of supersymmetry breaking in which the SUSY violating terms are generated below the scale where flavor effects are important.  In the final section we discuss a mechanism which takes this second approach of low energy supersymmetry breaking.  A promising feature of models with low energy supersymmetry breaking is that the new physics occurs within a well understood theory of gauge interactions and that it could possibly lead to experimental consequences.

\chapter{$\mu \rightarrow e + \gamma$}
\label{chap:calc}

\section{Importance of $\mu \rightarrow e + \gamma$}

Flavor changing processes have played an important role in the discovery of new physics. $K^0 - \bar{K}^0$ mixing experiments led to the development of the GIM mechanism \cite{gim} and gave us our first view of CP violation.  Another flavor changing process that could yield important results is the process $\mu \rightarrow e + \gamma$.

Individual lepton number is conserved in the standard model, so any signal for $\mu \rightarrow e + \gamma$ is evidence for new physics.  This could in fact be our first evidence of physics beyond the standard model if supersymmetry is not found at LEP II.  The process $\mu \rightarrow e + \gamma$ will yield much greater information, however, when it is combined with the measured values of supersymmetric particles, perhaps from LEP II, LHC or the NLC.  $\mu \rightarrow e + \gamma$ will give precision results for slepton mass splittings and mixing angles which will help us understand the biggest questions in the SSM:  the structure of flavor and the breaking of supersymmetry.  Since it is likely that the scale of new physics beyond the weak scale is above the reach of direct experiment, precision tests of the low energy parameters will be our only probe of high energy physics.

Although the SUSY contribution to flavor violation can in principle also be obtained from hadronic processes, leptonic processes provide a superior means.  Hadronic processes contain a standard model contribution which could dwarf the SUSY contribution.  Since we cannot accurately calculate the standard model contribution due to hadronic uncertainties, we could only measure the SUSY contribution if it is near the size of the standard model contribution.  In leptonic processes, any size of SUSY contribution can be measured, subject to experimental accuracy.  Furthermore, the process $\mu \rightarrow e + \gamma$ depends only on the supersymmetric parameters associated with charginos, neutralinos, and sleptons.  These are the parameters we should measure first and most accurately.

The current limit for the branching ratio of $\mu \rightarrow e + \gamma$ is $4.9 \times 10^{-11}$ \cite{pdb}.  The Mega Experiment has a proposal to measure the branching ratio with a sensitivity of $5 \times 10^{-13}$ \cite{mega}.  Recent calculations in the minimal $SU(5)$ SUSY GUT and $SO(10)$ SUSY GUT \cite{bh} show that the current $\mu \rightarrow e + \gamma$ branching ratio may be very near the experimental limits even in the absence of large new flavor effects.  In chapter~\ref{chap:prob} we show that we do expect large new flavor effects in realistic theories, so the current experimental limits already put strong constraints on the structure of flavor and the nature of supersymmetry breaking.  The extension of these limits could realistically see lepton violation, and in the event that it does not, this information in itself will give hints to the physics of flavor and supersymmetry breaking.

\section{Motivation for the Complete Calculation}

The first $\mu \rightarrow e + \gamma$ calculation was done in reference~\cite{ellis}.  Since then, the process has been studied by numerous authors, in particular \cite{calc}.  However, $\mu \rightarrow e + \gamma$ has not received the attention that has been given to the other flavor changing processes.  The full calculation has never been done.

If a positive result for $\mu \rightarrow e + \gamma$ is obtained, the value of the full calculation is clear.   However, even for order of magnitude estimates, current calculations are accurate only in specialized regions of parameter space.

On the basis of dimensionless couplings and dimensionful parameters, several neutralino and chargino diagrams are of the leading order.  However, calculations in the literature include only a fraction of these diagrams.  Additionally, there is a large variation in the size of dimensionless loop integrals which in some cases is the determining factor in which contribution dominates.  As a result, all diagrams should be considered even for order of magnitude calculations.

\section{Calculation}

The lepton flavor sector contains the Yukawa superpotential term of equation~\ref{eq:lepsup} and the SUSY violating terms of equation~\ref{eq:viol}.  We use the conventions defined in section~\ref{sec:ssm} for these equations.
\begin{equation}
\label{eq:lepsup}
W_{\rm lep} = l {\lambda}_{e}\bar{e}H_d
\end{equation}
\begin{equation}
\label{eq:viol}
{\cal{L}}_{\rm soft}= m_{l}^2 \tilde{l} \tilde{l}^{*} + m_{\bar{e}}^2 \tilde{\bar{e}} \tilde{\bar{e}}^{*} + l A_{e} {\lambda}_{e}\bar{e}H_d
\end{equation}

\begin{figure}
	\epsfysize 15cm
	\centerline{\epsffile{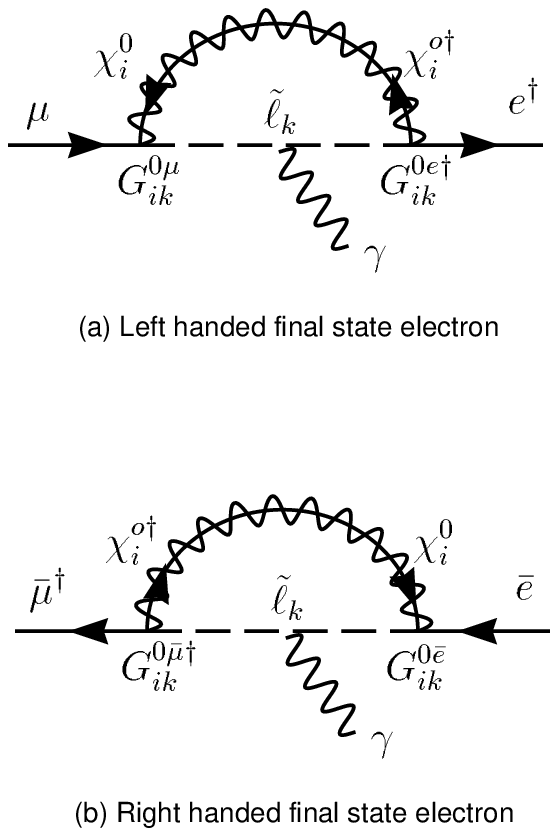}}
	\caption{{$\mu \rightarrow e+\gamma$} diagrams with a neutralino
	exchange and an external chirality flip.}
	\label{fig:ne}
\end{figure}

\begin{figure}
	\epsfysize 15cm
	\centerline{\epsffile{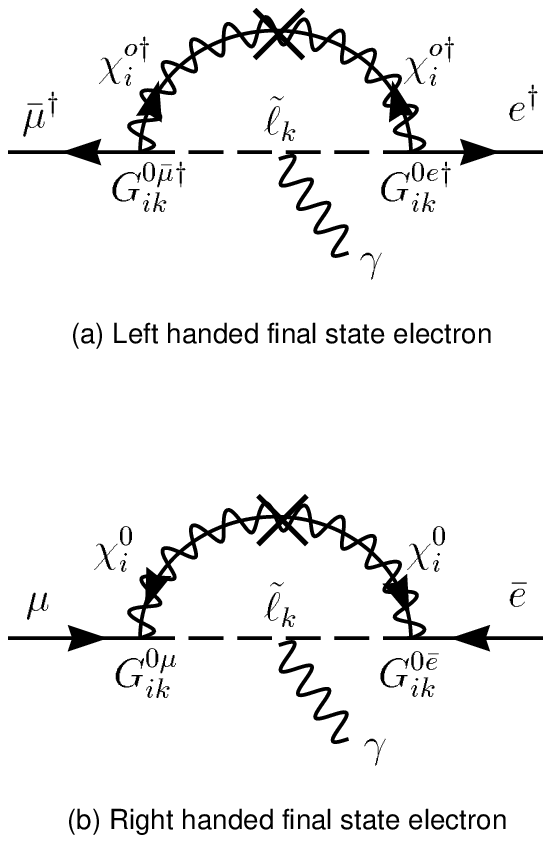}}
	\caption{{$\mu \rightarrow e+\gamma$} diagrams with a neutralino
	exchange and an internal chirality flip.}
	\label{fig:ni}
\end{figure}

\begin{figure}
	\epsfysize 15cm
	\centerline{\epsffile{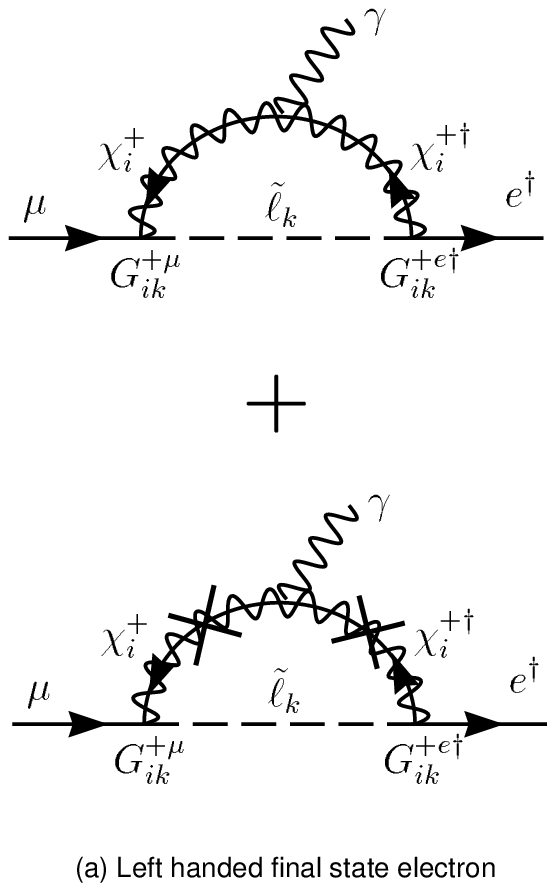}}
	\caption{{$\mu \rightarrow e+\gamma$} diagrams with a chargino
	exchange and an external chirality flip.}
	\label{fig:ce}
\end{figure}

\begin{figure}
	\epsfysize 15cm
	\centerline{\epsffile{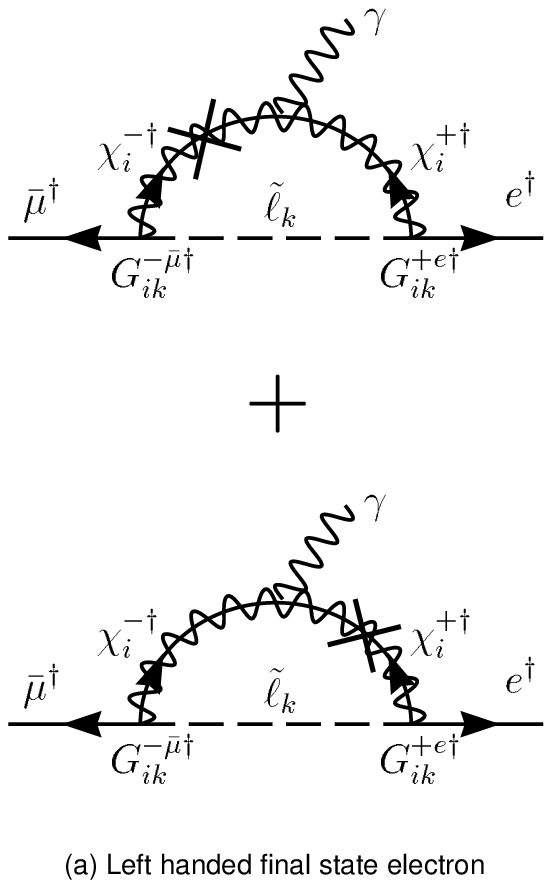}}
	\caption{{$\mu \rightarrow e+\gamma$} diagrams with a chargino
	exchange and an internal chirality flip.}
	\label{fig:ci}
\end{figure}

We will do a superfield redefinition using the $U(3)^2$ flavor rotation to make the Yukawa couplings diagonal so the external state leptons are in a mass eigenbasis.  The Yukawa couplings still contain a $U(1)^3$ symmetry corresponding to individual lepton number so any calculation which only includes the Yukawa and gauge couplings, the standard model contribution, will not give flavor violations.  The remaining terms of the flavor sector, the slepton masses and the A-terms, are in general non-diagonal and thus violate the $U(1)^3$ symmetry and give lepton flavor violation.
The one loop slepton contribution to $\mu \rightarrow e + \gamma$ consists of penguin diagrams where the internal states are a slepton and a neutralino or chargino.  We will consider all such diagrams with the assumption that the electron mass is zero (because it is much smaller than the muon mass).  As a result, there is no chargino diagram which couples to a final state $SU(2)$ singlet electron.  All relevant diagrams are shown in figures~\ref{fig:ne} through~\ref{fig:ci}.

The states in the loop are written in the lepton mass eigenbasis.  $\tilde{l}_{k}$ refers to a slepton, where $k$ is a flavor index running from one to six for selectrons and one to three for sneutrinos.  $\chi^0_i$ refers to one of the four neutralinos; $\chi^+_i$ refers to a chargino composed of the particles $\tilde{W}^+$ and $\tilde{\phi}_d$; and $\chi^-_i$ refers to a chargino composed of the particles $\tilde{W}^-$ and $\tilde{\phi}_u$.  The ``X'' that appears on the fermion propagators inside the loops does not refer to a single mass insertion, but a chirality flip of the propagator or, in other words, {\it every} odd number of mass insertions.  The Lagrangian conventions and mass matrices are given in appendices~\ref{app:con}~and~\ref{app:mass}.

All flavor violations are moved to the vertices, which we label $G$.  The superscripts tell which type of chargino or neutralino and which external lepton are in the coupling.  The subscripts label the chargino/neutralino ($i$) and the slepton ($k$).  These $G$ factors thus contain the matrix to rotate from the lepton mass basis to the slepton mass basis, and the matrix to rotate from the gaugino/Higgsino mass basis to the neutralino/chargino mass basis.  The couplings are given in appendix~\ref{app:coup}.


If the outgoing photon is on shell, then the operator form of the above diagrams can be decomposed into the two contributions of equation~\ref{eq:form} \cite{cl}.  One is the charge renormalization operator and the other is the anomalous magnetic moment.  By gauge invariance, flavor changing can only come from the magnetic moment operator.  

\begin{equation}
{\cal{L}}_{\rm loop} 
=
C_{1}(\psi \sigma^{\alpha} \psi^{\dagger}) A_{\alpha}
+
C_{2}(\bar{\psi}^{\dagger} \bar{\sigma}^{\alpha\beta} \psi^{\dagger}) F_{\alpha\beta}
\label{eq:form}
\end{equation}

Equation~\ref{eq:opform} gives the resulting flavor violating effective Lagrangian operators \cite{oper}.  As stated above, the $G$ factors give the flavor violating couplings of the muon or electron to the $k$ slepton eigenstate of mass $m_{k}$ and the $i$ chargino/neutralino eigenstate of mass $M_{i}$.  The loop functions $f$, $g$, $h$, and $j$ are given in appendix~\ref{sec:functions}.

\begin{equation}
\label{eq:opform}
{\cal{L}}_{\mu \rightarrow e+\gamma}^{eff} 
=
\frac{1}{2}\left\{ (\bar{\mu}^{\dagger} \bar{\sigma}^{\alpha\beta} e^{\dagger}) A_{l}+(\mu \sigma^{\alpha\beta} \bar{e}) A_{r} \right\} F_{\alpha \beta}
\end{equation}

\begin{eqnarray}
A_{l} 
=
\frac{e}{32\pi^{2}} & \left\{ G^{0 \mu}_{ik} G^{0e\dagger}_{ik} \frac{m_{\mu}}{m^{2}_{k}} f(\frac{M^{2}_{i}}{m^{2}_{k}}) + G^{0 \bar{\mu}\dagger}_{ik} G^{0e\dagger}_{ik} \frac{M_{i}}{m^{2}_{k}} h(\frac{M^{2}_{i}}{m^{2}_{k}}) \right. \nonumber \\ & \left. - G^{+ \mu}_{ik} G^{+e\dagger}_{ik} \frac{m_{\mu}}{m^{2}_{k}} g(\frac{M^{2}_{i}}{m^{2}_{k}}) - G^{- \bar{\mu}\dagger}_{ik} G^{+e\dagger}_{ik}\frac{M_{i}}{m^{2}_{k}} j(\frac{M^{2}_{i}}{m^{2}_{k}}) \right\} \\
A_{r} 
=
\frac{e}{32\pi^{2}} & \left\{ G^{0 \bar{\mu}\dagger}_{ik} G^{0\bar{e}}_{ik} \frac{m_{\mu}}{m^{2}_{k}} f(\frac{M^{2}_{i}}{m^{2}_{k}}) - G^{0 \mu}_{ik} G^{0\bar{e}}_{ik} \frac{M_{i}}{m^{2}_{k}} h(\frac{M^{2}_{i}}{m^{2}_{k}}) \right\}
\end{eqnarray}

\noindent The resulting $\mu \rightarrow e+ \gamma$ branching ratio is given in equation~\ref{eq:branch}.

\begin{equation}
BR(\mu \rightarrow e+\gamma)=\tau_{\mu} \frac{m^{3}_{\mu}}{16\pi} (|A_{l}|^{2}+|A_{r}|^{2})
\label{eq:branch}
\end{equation}

\section{Calculation Notes}

\subsection{Leading Graphs}
\label{sec:yuk}

The effective Lagrangian operator, equation~\ref{eq:opform}, contains a chirality flip between the lepton doublet and the lepton singlet.  This flip of the lepton can occur only through either a trilinear coupling in the slepton mass matrix, a Yukawa coupling at a Higgsino vertex, or a lepton mass insertion outside of the loop.  If the trilinear coupling is of order $M_{\rm weak}$, than the graphs including this term will dominate and cause serious flavor changing consequences.  We will instead assume approximate proportionality (as introduced in section~\ref{sec:fc}) in which the tri-linear couplings contain small dispersions from proportionality.  In this case, all $\mu \rightarrow e + \gamma$ contributions must contain a factor of a small Yukawa coupling; either through the A-term, Higgsino vertex, or external mass insertion.

Accompanying the Yukawa coupling factor will be a factor of the Higgs VEV, since the effective Lagrangian term also violates SU(2).  In the case of the A-term or the external mass insertion, this VEV factor is included with the Yukawa coupling.  In the case of the Higgsino vertex, the VEV factor will arise from the mixing between the Higgsino and the wino or bino, which is necessary to couple to the electron.

All graphs that contain a factor of the muon Yukawa and the Higgs VEV are potentially important, even in the case of a large SUSY scale.  We may neglect diagrams that contain more than one factor of the Yukawa coupling.

\subsection{Large Loop Functions}

\begin{figure}
	\label{fig:inv}
	\epsfysize 8cm
	\centerline{\epsffile{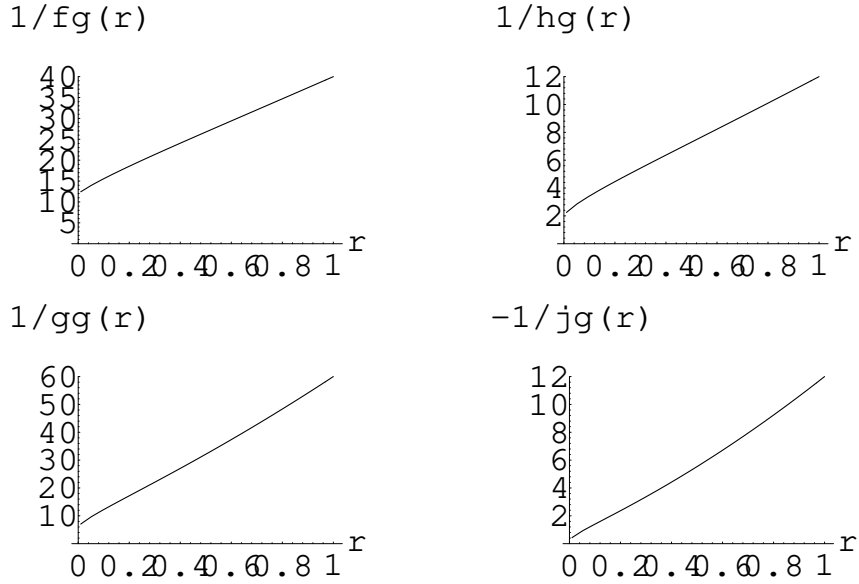}}
	\caption{Loop functions expanded for a small slepton mass difference.
	(Note that the inverse is plotted.)  The functions $f_g$ and $h_g$ 
	represent neutralino diagrams, and the function $g_g$ and $j_g$ 
	represent chargino diagrams.  $f_g$ and $g_g$ have an external 
	chirality flip, and $h_g$ and $j_g$ have an internal chirality flip.}
\end{figure}

To evaluate the flavor changing amplitude, the slepton, neutralino, and chargino masses and rotation matrices must be found, for which the necessary information is given in appendix~\ref{app:lag}.  In most cases, the slepton mass splitting and mixing angles are small parameters in which the amplitudes should be expanded.  The relevant quantity that determines the size of a loop contribution is in most cases not the loop functions $f$, $g$, $h$, and $j$, but the quantities defined in appendix~\ref{sec:functions}: $f_g$, $g_g$, $h_g$, and $j_g$, which result from an expansion in a small mass difference between the sleptons.  In figure~\ref{fig:inv}, we graph the inverse of these functions to illustrate the size variation among them.

The loop functions $h_g$ and $j_g$, which correspond to the diagrams which contain a chirality flip inside the loop, are significantly larger than the functions $f_g$ and $g_g$\footnote{The reason for this is that there is a smaller power of the momentum in the loop integral numerator of the chirality flip diagrams, giving a larger result.}.  In particular, the function $j_g$ diverges in the limit that the chargino mass goes to zero.  In many cases, the diagrams with internal chirality flips are dominant simply by virtue of the large dimensionless loop functions.

\subsection{Scaling the SUSY Masses}

Like all SUSY amplitudes, the amplitude for $\mu \rightarrow e + \gamma$ decreases as the SUSY scale increases.  The operator that generates the transition is of dimension five, and it contains one power of the fermion mass in the numerator, so the amplitude for the transition scales like $M^{-2}_{\rm SUSY}$.  If the SUSY scale is increased by a factor of $x$, the contour graph will remain unchanged if its axes are labeled by masses $x$ times larger and the values of the contours, the fractional mass splitting, represent values $x^{2}$ times larger.

We must of course clarify what is meant by increasing the SUSY scale.  This corresponds to increasing every element of the superpartners' mass matrices, making all new SUSY particles correspondingly heavier.  We can not actually do this scaling because some elements of the mass matrices come from the Higgs VEV, which we must to keep fixed.  Fortunately, the dependence on the Higgs VEV is not large because it is factored out in the lowest order approximation, as discussed in section~\ref{sec:yuk}.  Higher order effects are however still in the mass matrices.  If we scale only the parameters $M$, $m$, and $\mu$, then we will have the corresponding scaling of the mass splitting constraints, except in regions where the gaugino/Higgsino mass parameters are comparable to mass matrix elements arising from the Higgs VEV.  This region will have non-trivial scaling behavior.

The violation of the scaling is evident from examining  figures~\ref{fig:m50}~to~\ref{fig:m400} in chapter~\ref{chap:prob} which give the fractional mass splitting constraints on the sparticles as a function of SUSY parameter space.  There is little variation among the constraints from the range of parameter space corresponding to $M_{0}=50$ GeV, $\mu=-50$ GeV to $50$ GeV, and $m_{0}=100$ GeV.  However, there is a large variation among the constraints from the region $M_{0}=200$ GeV, $\mu=-200$ GeV to $200$ GeV, and $m_{0}=400$ GeV.  The effect of scaling violations is only dramatic for regions where there is a large cancellation and hence a sensitivity to the exact numerical results.  Outside of the excluded region of parameter space, scaling should give the correct order of magnitude in most cases.

\section{Summary}

The process $\mu \rightarrow e + \gamma$ is a sensitive probe of physics beyond the standard model.  Nonetheless, it has not been thoroughly studied -- previous branching ratio calculations omitted several important contributions.  In this paper, we present the complete $\mu \rightarrow e + \gamma$ calculation for the general supersymmetric standard model.

\newpage

\appendix

\chapter{Lagrangian}
\label{sec:lagrangian}
\label{app:lag}

\section{Conventions}
\label{app:con}

\paragraph\ We use the SUSY Lagrangian conventions of Wess and Bagger \cite{susy}, with two exceptions.  First, we use the dagger symbol to represent complex conjugation at all times.  The overbar symbol we use for the names of some of the standard model particles, for example, the quarks in the anti-fundamental representation.  Also, we will do an R rotation to make the gaugino couplings to the scalar superfields real, instead of imaginary as Wess and Bagger use.

Equation~\ref{eq:super} lists the superpotential.
\begin{equation}
\label{eq:super}
W = \mu \phi_{1} \phi_{2} + \lambda^{e}_{ik} \phi_{1} \bar{e}_{i} l_{k} + \lambda^{d}_{ik} \phi_{1} \bar{d}_{i} q_{k} + \lambda^{u}_{ik} \phi_{2} \bar{u}_{i} q_{k}
\end{equation}

Equation~\ref{eq:soft} lists the soft SUSY breaking terms \cite{il}.  The last term in the equation is the soft scalar masses.
\begin{eqnarray}
\label{eq:soft}
{\cal{L}}_{\rm soft} = - \left\{ B \mu \phi_{1} \phi_{2} + A^{e}_{ik} \lambda^{e}_{ik} \phi_{1} \tilde{\bar{e}}_{i} \tilde{l}_{k} + A^{d}_{ik} \lambda^{d}_{ik} \phi_{1} \tilde{\bar{d}}_{i} \tilde{q}_{k} + A^{u}_{ik} \lambda^{u}_{ik} \phi_{2} \tilde{\bar{u}}_{i} \tilde{q}_{k} + \right. \nonumber \\ \left. \frac{M_{1}}{2} \tilde{B} \tilde{B} + \frac{M_{2}}{2} \tilde{W} \tilde{W} + \frac{M_{3}}{2} \tilde{G} \tilde{G} + h.c. \right\} - \left\{ m^{2}_{ij} \phi_{i} \phi_{j} \right\}
\end{eqnarray}

We use the convention that for the product of the two Higgs doublets, the term that links the neutral components carries the same sign as the coefficient of the whole product.

We have the freedom to remove one more sign and still sample the entire parameter space.  We choose to make $\tan{\beta}$ always positive by making the sign of $B \mu$ negative.  This will effect the up and down Yukawa matrices, which are unimportant to the $\mu \rightarrow e+\gamma$ calculation.

We have made all the SUSY breaking parameters real because the effects of the actual phase must be very small from phenomenological considerations. 

\section{Mass Matrices}
\label{app:mass}

Equation~\ref{eq:nmass} gives the neutralino mass matrix \cite{hk}.
\begin{eqnarray}
\label{eq:nmass}
\begin{array}{ccc}
\left( \begin{array}{cccc}
\tilde{B} & \tilde{W}_{3} & \tilde{\phi}_{1}^0 & \tilde{\phi}_{2}^0 
\end{array} \right) & 
\left( \begin{array}{cccc}
M_{1} & 0 & -g_{1} \nu_{1} / \sqrt{2} & g_{1} \nu_{2} / \sqrt{2} \\
0 & M_{2} & g_{2} \nu_{1} / \sqrt{2} & -g_{2} \nu_{2} / \sqrt{2} \\
-g_{1} \nu_{1} / \sqrt{2} & g_{2} \nu_{1} / \sqrt{2} & 0 & \mu  \\
g_{1} \nu_{2} / \sqrt{2} & -g_{2} \nu_{2} / \sqrt{2} & \mu  & 0
\end{array} \right) &
\left( \begin{array}{c}
\tilde{B} \\ \tilde{W}_{3} \\ \tilde{\phi}_{1}^0 \\ \tilde{\phi}_{2}^0 
\end{array} \right)
\end{array}
\end{eqnarray}

Equation~\ref{eq:cmass} gives the chargino mass matrix \cite{hk}.
\begin{eqnarray}
\label{eq:cmass}
\begin{array}{ccc}
\left( \begin{array}{cc}
\tilde{W}^{+} & \tilde{\phi}_{2}^{+} 
\end{array} \right) & 
\left( \begin{array}{cc}
M_{2} & g_{2} \nu_{1} \\
g_{2} \nu_{1} & -\mu  \\
\end{array} \right) &
\left( \begin{array}{c}
\tilde{W}^{-} \\ \tilde{\phi}_{1}^{-} 
\end{array} \right)
\end{array}
\end{eqnarray}

For the leptons, we will work in a flavor basis where the scalar masses are diagonal.  The mass texture we assume in section~\ref{sec:mass} for our calculation is real and contains no mixing with the third family. Equation~\ref{eq:emass} gives the lepton mass matrix to the leading order in the mixing angles.

\begin{eqnarray}
\label{eq:emass}
\begin{array}{ccc}
\left( \begin{array}{cc}
\bar{e} & \bar{\mu}
\end{array} \right) & 
\left( \begin{array}{cc}
m_{e}+\theta_{l}\theta_{r}m_{\mu} & \theta_{l}m_{e}-\theta_{r}m_{\mu} \\
\theta_{r}m_{e}-\theta_{l}m_{\mu} & m_{\mu}+\theta_{l}\theta_{r}m_{e}
\end{array} \right) &
\left( \begin{array}{c}
e \\ \mu 
\end{array} \right)
\end{array}
\end{eqnarray}

Equation~\ref{eq:semass} gives the selectron mass matrix and equation~\ref{eq:snmass} gives the sneutrino mass matrix.  Here, we have dropped the $A$ terms proportional to the electron Yukawa and the second power of the mixing angles.  Note that the term $A^{*}$ refers to $A+\mu\tan{\beta}$.  Also, we have neglected the F-terms proportional to the Yukawa couplings squared.  Included in the diagonal mass squareds here is the D-term contribution from electroweak breaking.  This contribution, which is universal for particles of a given quantum number, is given in equations~\ref{eq:d1}~through~\ref{eq:d3} \cite{hk}.  

\begin{eqnarray}
\label{eq:semass}
\begin{array}{cc}
\left( \begin{array}{cccc}
m_{\tilde{e}}^2 & 0 & 0 & -A^{*}_{\bar{\mu} e} \theta_{l} m_{\mu} \\
0 & m_{\tilde{\mu}}^2 & -A^{*}_{\mu \bar{e}} \theta_{r} m_{\mu} & A^{*}_{\mu \bar{\mu}}m_{\mu} \\
0 & -A^{*}_{\mu \bar{e}} \theta_{r} m_{\mu} & m_{\tilde{\bar{e}}}^2 & 0  \\
-A^{*}_{\bar{\mu} e} \theta_{l} m_{\mu} & A^{*}_{\mu \bar{\mu}}m_{\mu} & 0  & m_{\tilde{\bar{\mu}}}^2
\end{array} \right) &
\left( \begin{array}{c}
\tilde{e} \\ \tilde{\mu} \\ \tilde{\bar{e}} \\ \tilde{\bar{\mu}} 
\end{array} \right)
\end{array}
\end{eqnarray}

\begin{eqnarray}
\label{eq:snmass}
\begin{array}{cc}
\left( \begin{array}{cc}
m_{\tilde{\nu}_{e}}^2 & 0 \\
0 & m_{\tilde{\nu}_{\mu}}^2 \\
\end{array} \right) &
\left( \begin{array}{c}
\tilde{\nu_{e}} \\ \tilde{\nu_{\mu}} 
\end{array} \right)
\end{array}
\end{eqnarray}

\begin{eqnarray}
\label{eq:d1}
m^{2}_{\tilde{e}} & = & m^{2}_{\tilde{l},soft}+ \left( \frac{-g^{2}_{1}+g^{2}_{2}}{4} \right) (v^{2}_{2}-v^{2}_{1}) \\
m^{2}_{\tilde{\nu}} & = & m^{2}_{\tilde{l},soft}+ \left( \frac{-g^{2}_{1}-g^{2}_{2}}{4} \right) (v^{2}_{2}-v^{2}_{1}) \\
m^{2}_{\tilde{\bar{e}}} & = & m^{2}_{\tilde{\bar{e}},soft}+ \left( \frac{g^{2}_{1}}{2} \right) (v^{2}_{2}-v^{2}_{1})
\label{eq:d3}
\end{eqnarray}

\section{Couplings}
\label{app:coup}

We now give information for the Lagrangian couplings labeled $G_{ik}^{cf}$ in the text.  The lower indices $i$ and $k$ refer to the mass eigenstates for the charginos/neutralinos and the sleptons in the vertex, respectively.  The upper index $c$ represents the charge of the neutralino or chargino, and $f$ represents the incoming fermion. 

Below, we will display the couplings in the gauge/lepton mass eigenbasis.  The couplings must be rotated to the neutralino/chargino mass basis with the use of the proper mass matrices.

\subsection{Gauginos}

The gaugino couplings are of course the same for all particles with a given quantum number, and they are diagonal in family space.  Below we display the coupling of the lepton to the complex conjugate of the slepton.

\begin{eqnarray}
G^{0 e}_{W \tilde{e}} & = & -g_{2} / \sqrt{2} \\
G^{+ e}_{W \tilde{\nu}} & = & g_{2} \\
G^{0 e}_{B \tilde{e}} & = & -g_{1} / \sqrt{2} \\
G^{0 \bar{e}}_{B \tilde{\bar{e}}} & = & \sqrt{2} g_{1}
\end{eqnarray}

\subsection{Higgsinos}
We only use the Higgsino couplings for the muon because it is much larger than the electron.  Here we display the coupling of the muon to the slepton. 

\begin{eqnarray}
G^{0 \mu}_{\phi \tilde{\bar{\mu}}} & = & -m_{\mu}/v_{1} \\
G^{0 \bar{\mu}}_{\phi \tilde{\mu}} & = & -m_{\mu}/v_{1} \\
G^{- \bar{\mu}}_{\phi \tilde{\nu}} & = & +m_{\mu}/v_{1}
\end{eqnarray}

\chapter{Special Function Definitions}
\label{sec:functions}

\section{$\mu \rightarrow e + \gamma$ Loop Functions}

\paragraph\ Equations~\ref{eq:loop1}~through~\ref{eq:loop4} give the loop functions for $\mu \rightarrow e + \gamma$.
\begin{eqnarray}
\label{eq:loop1}
f(r) & = & \frac{1}{12(1-r)^4} \left( 2r^3+3r^2-6r+1-6r^2\log{r} \right) \\
g(r) & = & \frac{1}{12(1-r)^4} \left( r^3-6r^2+3r+2+6r\log{r} \right) \\
h(r) & = & \frac{1}{2(1-r)^3} \left( -r^2+1+2r\log{r} \right) \\
j(r) & = & \frac{1}{2(1-r)^3} \left( r^2-4r+3+2\log{r} \right)
\label{eq:loop4}
\end{eqnarray}

\section{$\mu \rightarrow e + \gamma$ Amplitude Functions}

In section~\ref{sec:mass}, we calculated the transition amplitude in the context of a particular theory of lepton masses.  In this appendix we give the necessary functions for equation~\ref{eq:gamp}.  We use modified loop functions which are defined below.  The argument of these loop function is $r_{pk}=M_{k}^{2}/m_{p}^2$ where $k$ represents the chargino or neutralino, and $p$ represents the slepton. 

The $U$ matrices rotate the gaugino/higgsino interaction basis into the neutralino/chargino mass basis.  $U^{0}$ is for the neutralinos; $U^{+}$ is for the charginos $\tilde{W}^{+}$ and $\tilde{H}_{u}^{+}$; and $U^{-}$ is for the charginos $\tilde{W}^{-}$ and $\tilde{H}_{d}^{-}$.
\begin{eqnarray}
X_{l} & = & X_{lf} + X_{lh} + X_{lg} + X_{lj} \\
X_{r} & = & X_{rf} + X_{rh} \\
X_{lf} & = & \frac{1}{2} \left( U^{0}_{Wk} + \frac{g_{1}}{g_{2}} U^{0}_{Bk} \right)^{2} f_{g}(r_{ek}) \\
X_{rf} & = & -2 \left( \frac{g_{1}}{g_{2}} U^{0}_{Bk} \right)^{2} f_{g}(r_{\bar{e}k}) \\
X_{lh} & = & \frac{(A+\mu \tan{\beta}) M}{m^{2}_{\tilde{e}}} \left( \frac{g_{1}}{g_{2}} U^{0}_{Bk} \right) \left( U^{0}_{Wk} + \frac{g_{1}}{g_{2}} U^{0}_{Bk} \right) h_{k}(r_{ek},r_{\bar{e}k}) \nonumber \\
& & - \frac{M}{\sqrt{2} g_{2} v_{1}} \left( U^{0}_{H k} \right) \left( U^{0}_{Wk} + \frac{g_{1}}{g_{2}} U^{0}_{Bk} \right) h_{g}(r_{ek}) \nonumber \\
& & + \left( \frac{m^{4}_{\tilde{e}}}{\delta m^{2}_{\tilde{e}}} \right) \frac{\delta A_{\bar{\mu} e} M}{m^{2}_{\tilde{e}}-m^{2}_{\tilde{\bar{e}}}} \left( \frac{g_{1}}{g_{2}} U^{0}_{Bk} \right) \left( U^{0}_{Wk} + \frac{g_{1}}{g_{2}} U^{0}_{Bk} \right) \left[ \frac{h(r_{ek})}{m^{2}_{\tilde{e}}} - \frac{h(r_{\bar{e}k})}{m^{2}_{\tilde{\bar{e}}}} \right] \\
X_{rh} & = & \frac{(A+\mu \tan{\beta}) M}{m^{2}_{\tilde{\bar{e}}}} \left( U^{0}_{Wk} + \frac{g_{1}}{g_{2}} U^{0}_{Bk} \right) \left( \frac{g_{1}}{g_{2}} U^{0}_{Bk} \right) h_{k}(r_{\bar{e}k},r_{ek}) \nonumber \\
& & + \frac{\sqrt{2} M}{g_{2} v_{1}} \left( U^{0}_{H k} \right) \left( \frac{g_{1}}{g_{2}} U^{0}_{Bk} \right)^{2} h_{g}(r_{\bar{e}k}) \nonumber \\
& & + \left( \frac{m^{4}_{\tilde{\bar{e}}}}{\delta m^{2}_{\tilde{\bar{e}}}} \right) \frac{\delta A_{\mu \bar{e}} M}{m^{2}_{\tilde{\bar{e}}} - m^{2}_{\tilde{\bar{e}}}} \left( U^{0}_{Wk} + \frac{g_{1}}{g_{2}} U^{0}_{Bk} \right) \left( \frac{g_{1}}{g_{2}} U^{0}_{Bk} \right) \left[ \frac{h(r_{\bar{e}k})}{m^{2}_{\tilde{\bar{e}}}} - \frac{h(r_{ek})}{m^{2}_{\tilde{e}}} \right] \\
X_{lg} & = & -\left( \frac{m^{4}_{\tilde{e}}}{m^{4}_{\tilde{\nu}}} \right) \left( U^{+}_{Wk} \right)^{2} g_{g}(r_{\bar{\nu}k}) \\
X_{lj} & = & \left( \frac{m^{4}_{\tilde{e}}}{m^{4}_{\tilde{\nu}}} \right) \frac{M}{g_{2} v_{1}} \left( U^{-}_{H k} \right) \left( U^{+}_{Wk} \right) j_{g}(r_{\bar{e}k})
\end{eqnarray}

Although we set $A=0$ in the main text, we include the $A$ dependence here in the appendix.  Equations~\ref{eq:a1}~and~\ref{eq:a2} give definitions for the nonuniversality of the $A$ terms used above.
\begin{eqnarray}
\label{eq:a1}
\delta A_{\bar{\mu} e} = A_{\bar{\mu} e} - A_{\bar{\mu} \mu} \\
\delta A_{\bar{e} \mu} = A_{\bar{e} \mu} - A_{\bar{\mu} \mu}
\label{eq:a2}
\end{eqnarray}

For our original functions $f$, $g$, $h$, and $j$, we have two modifications that result from our expansion in the inter-family mass difference.  Equation~\ref{eq:gim} defines the $g$ subscript, and equation~\ref{eq:kim} defines the $k$ subscript.  $Z$ represents any of the four functions $f$, $g$, $h$, or $j$.
\begin{eqnarray}
\label{eq:gim}
Z_{g} \left( \frac{M^{2}}{m^{2}} \right) & \equiv & m^4 \frac{d}{dm^2} \left\{ \frac{1}{m^2} Z \left( \frac{M^2}{m^2} \right) \right\} \\
Z_{k} \left( \frac{M^{2}}{m^{2}_{a}} , \frac{M^{2}}{m^{2}_{b}} \right) & \equiv & m^6_{a} \frac{d}{dm^2_{a}} \left\{ \frac{1}{m^{2}_{a} - m^{2}_{b}} \left[ \frac{1}{m^{2}_{a}} Z \left( \frac{M^2}{m^{2}_{a}} \right) - \frac{1}{m^{2}_{b}} Z \left( \frac{M^2}{m^{2}_{b}} \right) \right] \right\}
\label{eq:kim}
\end{eqnarray}

\section{$K^{0}-\bar{K}^{0}$ Functions}

\paragraph\  Equations~\ref{eq:kk1}~and~\ref{eq:kk2} are the loop functions from the text in terms of the functions $f_{6}$ and $\tilde{f_{6}}$ of Hagelin et. al. \cite{hag}, shown in equations~\ref{eq:hag1}~and~\ref{eq:hag2}.

\begin{eqnarray}
\label{eq:kk1}
f_{1}(r) & = & -66 \tilde{f_{6}}(r) - 24 r f_{6}(r) \\
\label{eq:kk2}
f_{2}(r) & = & \left\{ -36-24 \left( \frac{m_{K}}{M_{s}+m_{d}} \right) ^2 \right\} \tilde{f_{6}}(r) + \nonumber
\\       &   & \left\{ -72+384 \left( \frac{m_{K}}{m_{s}+m_{d}} \right) ^2 \right\} r f_{6}(r) \\
\label{eq:hag1}
f_{6}(r) & = & \frac{1}{6(1-r)^5} \left( -r^3+9r^2+9r-17-18r\log{r}-6r\log{r} \right) \\
\label{eq:hag2}
\tilde{f_{6}}(r) & = & \frac{1}{3(1-r)^5} \left( r^3+9r^2-9r-1-6r^2\log{r}-6r\log{r} \right)
\end{eqnarray}

\chapter{$\mu \rightarrow e+\gamma$: $\tan{\beta}$ and $A$ dependence}
\label{sec:tanBA}
\label{app:Atan}

\paragraph\ For completeness, we give here the dependence of the lepton mass splitting on the parameters $\tan{\beta}$ and $A$.  One place these parameters appear is in graphs with a chirality flip in which the muon couples to a gaugino. Here, the flip between the left-handed and right-handed particles happens in the slepton mass matrix through a term proportional to  $A+ \mu \tan{\beta}$.  Another place these parameters appear is in the chirality flip graphs where the muon couples to the higgsino. The product of the muon Yukawa and the mixing between the down higgsino and the wino gives a term proportional to $\tan{\beta}$, once $m_{\mu}$ is factored out.  The contribution from these parameters is enhanced by the fact that they have the larger chirality flip loop functions as coefficients.
 
\begin{figure}
	\epsfysize 8.5cm
	\centerline{\epsffile{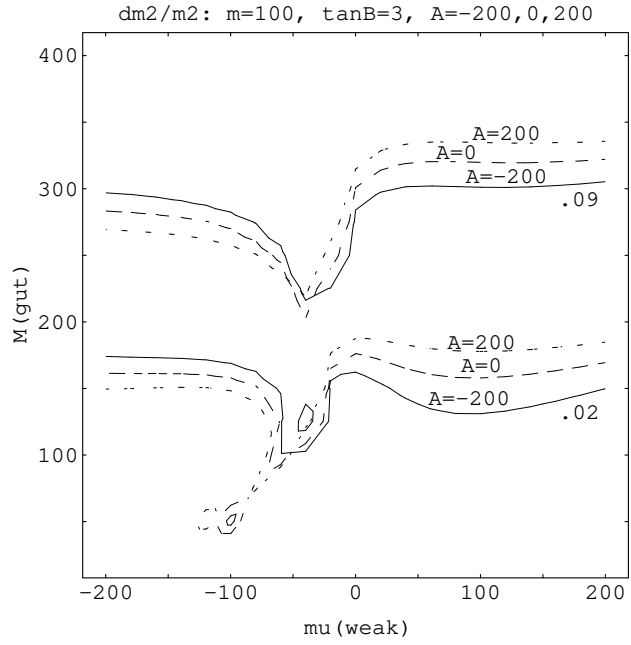}}
	\caption{Slepton mass splitting constraints from
		 three values of $A$}
	\label{fig:Agraph}
\end{figure}

Figure~\ref{fig:Agraph} shows the fractional lepton mass splitting at the GUT scale as a function of $\mu$ and $M_{0}$ for $m_{0}=100$, $\tan{\beta}=3$  and $A=-200,0,200$.  A particular value of A increases or decreases the constraint on the splitting depending on the relative sign with $\mu$.  Varying A does not give a large qualitative effect on the mass splitting.

\begin{figure}
	\epsfysize 8.5cm
	\centerline{\epsffile{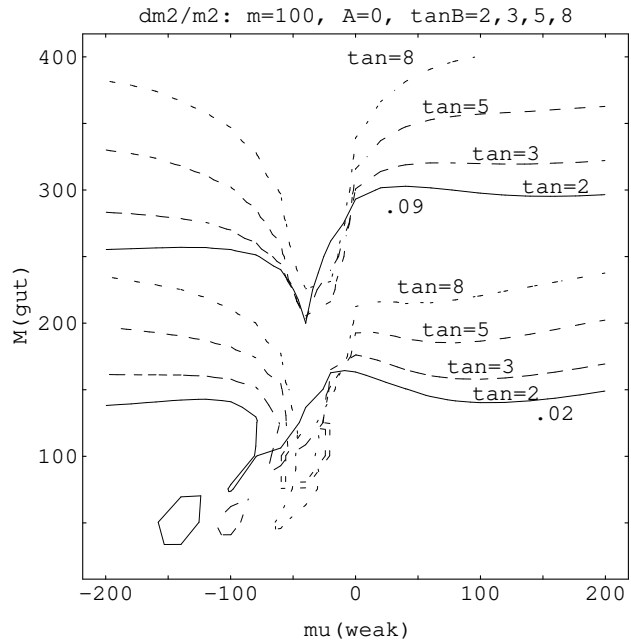}}
	\caption{Slepton mass splitting constraints from
		 four values of $\tan{\beta}$}
	\label{fig:tanBgraph}
\end{figure}

\begin{figure}
	\epsfysize 8.5cm
	\centerline{\epsffile{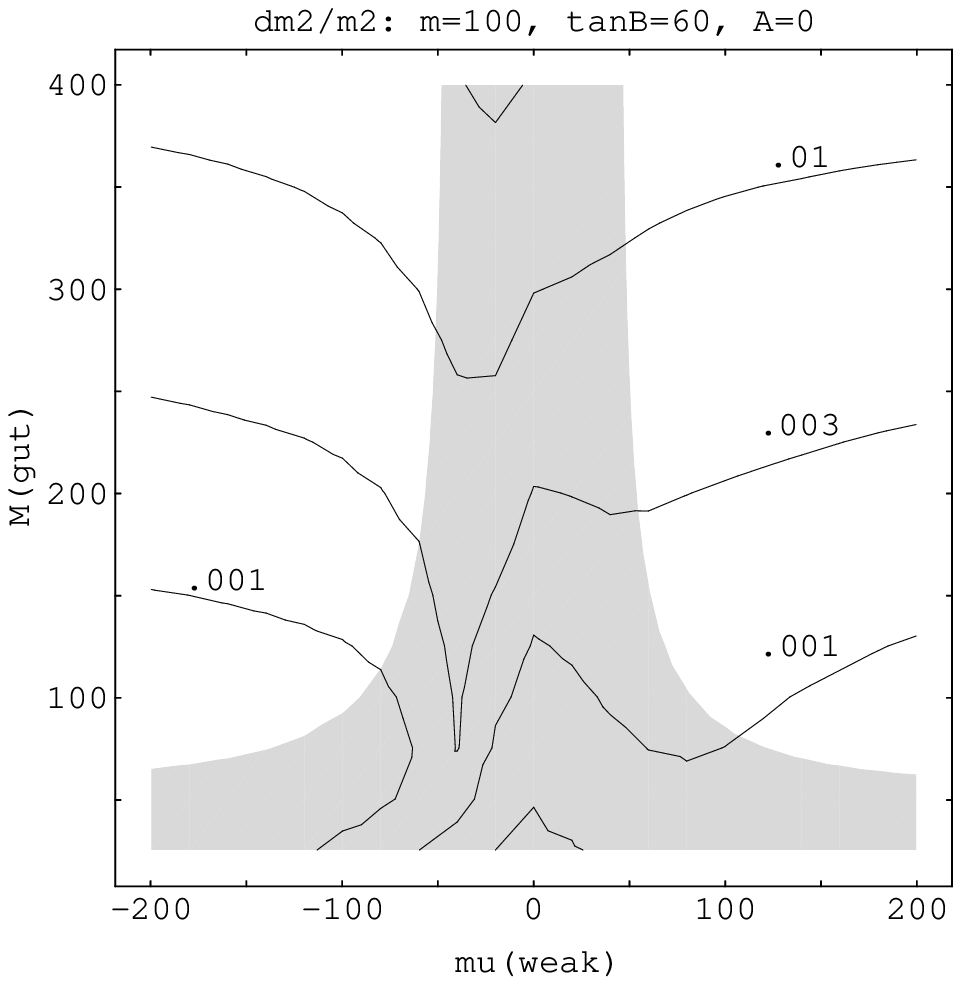}}
	\caption{Slepton mass splitting contraints from
		 $\tan{\beta}=60$}
	\label{fig:lartanBgraph}
\end{figure}

Figure~\ref{fig:tanBgraph} shows the fractional lepton mass splitting at the GUT scale as a function of $\mu$ and $M_{0}$ for $m_{0}=100$, $A=0$, and $\tan{\beta}=2,3,5,8$.  This dependence is strong, so that for larger values of $\tan{\beta}$, the mass splitting constraint scales almost linearly.  Figure~\ref{fig:lartanBgraph} shows the mass splitting for $\tan{\beta}=60$.  

\end{document}